\documentclass[namedreferences]{SolarPhysics}
\usepackage[optionalrh]{spr-sola-addons} 
\usepackage{graphicx}                    
\usepackage{amssymb}                    
\usepackage{color}                       
\usepackage{url}                         

\usepackage{solaheader}	
\newcommand{\solareceiveddate}{29 June 2010} 
\newcommand{\solaaccepteddate}{23 September 2010}

\begin{document}

\begin{article}

\begin{opening}

\title{Search for Rapid Changes in the Visible-Light Corona during the 21~June 2001 Total Solar Eclipse}

%
\author{P.~\surname{Rudawy}$^{1}$\sep
        K.J.H.~\surname{Phillips}$^{2}$\sep
        A.~\surname{Buczylko}$^1$\sep
        D.R.~\surname{Williams}$^2$\sep\\
        F.P.~\surname{Keenan}$^3$\sep
       }

%
\runningauthor{P. Rudawy {\it et al.}}
\runningtitle{Search for Rapid Changes in the Corona}

%
  \institute{$^{1}$ Astronomical Institute, University of Wroc{\l}aw, ul. Kopernika 11, 51-622 Wroc{\l}aw, Poland
                     email: \url{rudawy@astro.uni.wroc.pl,buczylko@astro.uni.wroc.pl} \\
             $^{2}$ Mullard Space Science Laboratory, University College London, Holmbury St Mary, Dorking, Surrey RH5 6NT, United Kingdom
                     email: \url{kjhp@mssl.ucl.ac.uk, drw@mssl.ucl.ac.uk} \\
             $^{3}$ Astrophysics Research Centre, School of Mathematics and  Physics, Queen's University Belfast, Belfast BT7 1NN, N. Ireland, United Kingdom email: \url{F.Keenan@qub.ac.uk}
             }

\begin{abstract}
 Some 8000 images obtained with the SECIS fast-frame CCD camera instrument located at Lusaka, Zambia, during the total eclipse of 21~June 2001 have been analyzed to search for short-period oscillations in intensity that could be a signature of solar coronal heating mechanisms by MHD wave dissipation. Images were taken in white-light and Fe~{\sc xiv} green-line (5303~\AA) channels  over 205~seconds (frame rate 39 s$^{-1}$), approximately the length of eclipse totality at this location, with a pixel size of four~arcseconds square. The data are of considerably better quality than were obtained during the 11~August 1999 total eclipse, observed by us (Rudawy {\it et al.}: {\it Astron. Astrophys.} {\bf 416,} 1179, 2004), in that the images are much better exposed and enhancements in the drive system of the heliostat used gave a much improved image stability. Classical Fourier and wavelet techniques have been used to analyze the emission at 29518 locations, of which 10714 had emission at reasonably high levels, searching for periodic fluctuations with periods in the range 0.1--17~seconds (frequencies 0.06--10~Hz). While a number of possible periodicities were apparent in the wavelet analysis, none of the spatially and time-limited periodicities in the local brightness curves was found to be physically important.  This implies that the pervasive Alfv\'en wave-like phenomena (Tomczyk {\it et al.}: {\it Science} {\bf 317,} 1192, 2007) using polarimetric observations with the CoMP instrument do not give rise to significant oscillatory intensity fluctuations.

\end{abstract}
\end{opening}

%
\section{Introduction}


Searches for the specific agent involved in the heating of the solar corona have so far failed to decide definitively between the two chief candidate mechanisms. The first is a ``DC" model in which the heating occurs by myriads of small reconnection events (``nanoflares") in flux tubes as a result of footpoint shuffling by photospheric motions \cite{lev74,par88}. Alternatively, in the ``AC" model the heating is a result of dissipation of Alfv\'en or magnetohydrodynamic (MHD) waves in loop structures \cite{hol82,kud99,por94} or more likely sub-resolution strands making up what are currently identified as loops (in, {\it e.g.}, TRACE images) \cite{car97,kli06}. Some important developments have been made in recent years that perhaps suggest that wave heating is at least important if not dominant. First, there is some evidence that the amount of energy available from integrating frequency distributions of observed small-scale phenomena such as tiny X-ray or EUV flares \cite{kru98,par00,asc04} falls short by a factor $\approx$~three of the required energy flux of the quiet solar corona ($3\times 10^{5}$ erg~cm$^{-2}$~s$^{-1}$). Secondly, direct observations of Alfv\'en waves have been made in the solar chromosphere  using the {\it Hinode Solar Optical Telescope} \cite{dep07} and corona using a high-resolution ground-based coronagraph \cite{tom07} respectively. The waves observed by \inlinecite{dep07} in particular have sufficient energy to drive the solar wind and account for a large proportion of the solar corona's observed energy.

Observations during total solar eclipses have an important bearing on the coronal-heating problem, since fast-cadence imaging is possible so that heating by short-period ($\approx 1$~second) MHD waves, suggested to be important by \inlinecite{por94}, can be investigated. Imaging cadences from spacecraft, even from the {\it Atmospheric Imaging Assembly} (AIA) on the recently launched {\it Solar Dynamics Observatory}, are still insufficient to search for such periods. Thus, weak periodic modulations of the intensity of the coronal Fe~{\sc xiv} green line (5303~\AA)  during eclipses in the 1980s by \inlinecite{pas87} have been reported, with better evidence in the period range 1--1.3~seconds (frequencies 0.75--1.0~Hz) from a more sophisticated CCD camera instrument during the 1999 eclipse \cite{rus00,pas02}. \inlinecite{sin09} analyzed their 2009 eclipse data and report on significant oscillatory power with periods of between 20 and 27~seconds in loop structures observed in the Fe~{\sc xiv} green line and the lower-temperature Fe~{\sc x} red line at the boundary of an active region.

In a previous paper \cite{rud04}, a search for fast modulations in the coronal emission measured with a CCD camera instrument during the total solar eclipse of 11~August 1999 was reported. The instrument, the {\it Solar Eclipse Coronal Imaging System}  (SECIS: \opencite{phi00}), consisted of a 300~mm heliostat, a horizontally mounted 200~mm reflecting telescope and heliostat, and a pair of fast-frame CCD cameras. During eclipse totality, these cameras each captured some 6364 images of the corona with each camera at a rate of 44~s$^{-1}$ in white light and in the green line. Some 6390 individual pixels and small areas were analyzed within the coronal features visible using classical Fourier techniques: some 17 cases of peaks in the power spectra were observed having maxima $>4\sigma$ above the background in the period range 0.1--1~seconds (1--10~Hz). However, application of rigorous statistical tests showed that none of these was significant. An earlier wavelet analysis \cite{wil01} of the same data shows evidence of a travelling wave moving down the leg of a small loop, with periodicities of six seconds (0.16~Hz).

Here we report on a search made with an enhanced version of the SECIS instrumentation during the total eclipse of 21~June 2001 with a longer totality duration (194~seconds) observed under near-ideal conditions (cloudless skies with high atmospheric transparency) from the campus of the State University of Zambia at Lusaka, Zambia. As will be shown, the quality of images was much improved over those from our instrument during the 1999 eclipse.

In making a detailed analysis of the 2001 eclipse results, we took into account the striking results from the {\it Coronal Multi-Channel Polarimeter} (CoMP) instrument at the National Solar Observatory, New Mexico, reported by Tomczyk and co-workers \cite{tom07,tom09}. With a tunable filter centred on the near-infrared line of Fe~{\sc xiii} at 10747~\AA, they find the line's central wavelength and line-of-sight velocity and Stokes parameters from which the degree of linear polarization and direction of the magnetic field can be derived. Movies of the velocity images in particular reveal large numbers of quasi-periodic fluctuations, especially along the loops of an active region system moving out from the solar limb with phase speeds of about $2000$~km s$^{-1}$ and velocity amplitudes of about 0.5~km~s$^{-1}$. These fluctuations are interpreted by \inlinecite{tom07} as Alfv{\'e}n waves. No intensity fluctuations were observed above a limit of 0.3\%. \inlinecite{doo08} have interpreted these velocity fluctuations as fast magneto-acoustic kink waves which oscillate with the kink frequency corresponding to a phase speed of $\approx 2000$~km s$^{-1}$, although \inlinecite{tom09} have defended their original interpretation as involving fewer assumptions.

With such a high-quality data set as the 2001 SECIS observations, we are in a position to say much more definitely than from our 1999 eclipse data whether there are any periodic intensity fluctuations in the green-line images that could be signatures for coronal heating processes. This article describes the instrument set-up used during the 2001 eclipse (Section~2), the coalignment of images and corrections for motions (Section~3), photometric analysis of the image data (Section~4), and results and conclusions (Section~5).

\section{SECIS Instrument during the 2001 Eclipse}

\subsection{Instrument and Alignment}

\begin{figure}
\centerline{\includegraphics[width=0.7\textwidth,clip=,angle=0]{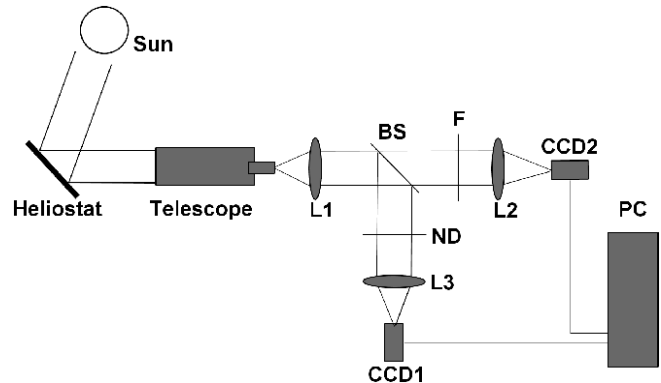}}
\caption{The layout of the SECIS instrument and heliostat during the 21~June 2001 eclipse. The principal optical elements include the heliostat, telescope, the three collimator lenses (L1, L2, L3), the beam splitter (BS), the green-line filter (F), and neutral-density filter (ND). The CCD cameras are CCD1 (white-light channel) and CCD2 (green-line channel), and PC is the computer and camera interfaces.}\label{instrument_layout}
\end{figure}

The SECIS instrument set-up was refined from the one used for the 1999 eclipse, described by \inlinecite{phi00}. Figure~\ref{instrument_layout} shows the instrument layout during the 2001 eclipse. The optical elements in this layout were a 200~mm, $f/10$ Meade Schmidt--Cassegrain reflecting telescope to provide an image of the eclipsed Sun, a collimating lens, a pellicle beam splitter to give two optical paths, a straight-through path (90\% transmission) and a reflected path (10\% transmission). Rays along the straight-through path were incident on a narrow-band filter centred on the 5303~\AA\ green line and a lens giving a focused image on one of the CCD cameras (CCD2). Rays reflected by the beam splitter were incident on a lens giving a focused image on the other CCD camera (CCD1). A neutral-density filter (transmission 6.3\%) was inserted in the second path. All optical components were mounted horizontally on a rigid aluminium optical bench resting on four feet, and enclosed in a light-tight aluminium box. Micrometers were used to adjust the focusing of each of the lenses in front of the cameras. Sunlight was reflected into a horizontal beam using the heliostat, also used during the 1999 eclipse, but with the drive motors enclosed in dust-proof metal boxes to exclude foreign particles that are now thought to have contributed to irregular tracking during the 1999 eclipse (these were compensated for during the analysis of the results obtained then: see \opencite{rud04}).

The SECIS cameras manufactured for our earlier expeditions to the 1998 and 1999 eclipses \cite{phi00} by EEV (Chelmsford, UK: renamed E2V), each have a monochrome $512 \times 512$ pixel frame transfer CCD with square pixels of dimension $15\mu {\rm m} \times 15\mu {\rm m}$. At a common trigger pulse from the control electronics, both cameras capture images with precise synchronization. The images are digitized to 12 bits, although with the two least significant bits judged to be noise this is effectively 10 bits. There is therefore a dynamic range of 1 to 1023. This is not particularly large, so we chose the transmission of the filters in the system with some care in order that images were neither over-exposed nor under-exposed. There was a very slight degrading of the cameras' performance since the 1999 eclipse, mainly near the edges of the sensitive area, but using flat-field and dark-current exposures taken after the eclipse these effects were largely removed.

The computer system operating the cameras is an adapted personal computer, provided by the Carr Crouch Computer Company (Maidenhead, UK), with specially written software. The system takes the synchronized video streams from each camera and reconstitutes video images. It has dual Pentium~II processors and 128~MB of memory. The video data are packed and stored on two circular buffer sets in main memory; when the memory is full, half-images from each camera are transferred to one of four 9-GB hard drives. Before a sequence of data is acquired, the user must select a total number of images which are pre-allocated in memory, a frame rate, exposure time and a short delay time to enable one image to be correctly captured before another is taken.

\subsection{Green-Line Filter}\label{Section_2.2}

\begin{figure}
\centerline{\includegraphics[width=.7\textwidth,clip=,angle=0]{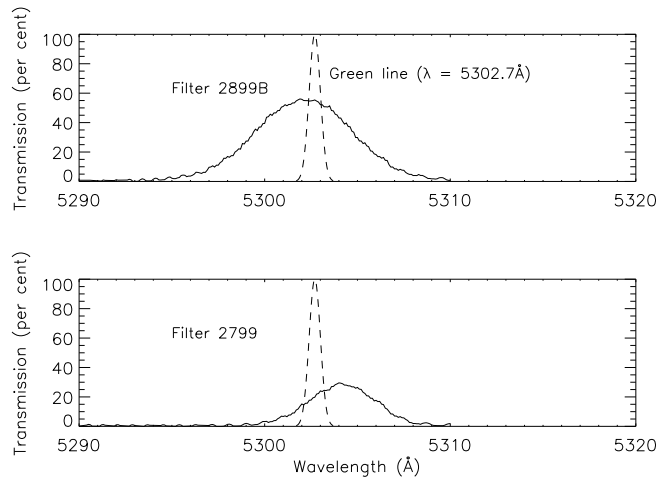}}
\caption{The measured transmission of green-line filters used in eclipses. Upper panel: Filter 2899B, used in the 2001 eclipse discussed in this paper.  Lower panel: Filter 2799, used in the 1999 eclipse. The Fe~{\sc xiv} green line theoretical profile, with thermal Doppler broadening (FWHM = 0.71~\AA) and central wavelength in air equal to 5302.7~\AA\ (see text), is also shown.}\label{filter_trans}
\end{figure}

We had available three green-line interference filters (serial numbers 2799, 2899A, 2899B) manufactured by Barr Associates, Inc. (Westford, Mass., USA) for our 1998 and 1999 eclipse expeditions. Comparison of the manufacturer's transmission curves led us to use filter number 2799 for the 1999 eclipse. However, subsequent measurements of the transmissions of all three filters with a spectrophotometer at University College London in January 2001 (largely made to check that the filters were still on-band) gave significantly different results from the original data; it is possible that these were due to changes in the nearly three years since their manufacture. As a result of this, we chose to use filter number 2899B for the 2001 eclipse. UCL measurements  of the two filters are shown in Figure~\ref{filter_trans}, where the peak transmission of filter number 2899B is nearly a factor two higher than that of filter number 2799 used in 1999, although the wavelengths of the peaks are 2~\AA\ apart. The width (FWHM) of the bandpass of filter number 2899B is 6.0~\AA, while that of filter number 2799 is 4.5~\AA. Subsequent measurements at Tatranska Lomnica Observatory in Slovakia of these same filters (which are being used on the coronagraph there) broadly confirm the UCL measurements.

The exact wavelength of the Fe~{\sc xiv} green line (transition $3s^2 3p\,^2P_{1/2} - 3s^2 3p\,^2P_{3/2}$) is of considerable importance in this respect since the peak transmission wavelength of the filter used must lie near it. This will be the subject of a discussion in a future work but in the meantime it can be said that, a mean of values published by \inlinecite{ryb86,kel87} and an analysis by R.J. Thomas (private communication, 2001) of extreme ultraviolet line wavelengths from the high-precision {\it Solar Extreme-ultraviolet Rocket Telescope and Spectrograph} (SERTS: \opencite{tho94}) gives the {\it in-vacuo} wavelength as 5304.2~\AA. With refractive index $[n]$ for air at a temperature and pressure approximately equal to those at the 2001 eclipse site given by $(n-1) \times 10^{-6} = 278$ \cite{all73}, the green-line wavelength in air is 5302.7~\AA. The thermal Doppler width (FWHM) of the line is given by $\Delta \lambda = 1.665 \times (\lambda/c)\times ({2 k_B T}/{m_{\rm Fe}})^{1/2}$,
where $k_B =$ Boltzmann's constant, $c =$ the velocity of light, and $m_{\rm Fe} = $ the mass of the Fe$^{+13}$ ion. For the temperature of maximum contribution function of the 5303~\AA\ line ($T=2$~MK), $\Delta \lambda = 0.71$~\AA. The line profile is over-plotted on each of the filter transmissions shown in Figure~\ref{filter_trans}. It may be seen that the transmission for the wavelength of the green line for filter 2799, used in the 1999 eclipse, is 20\%, but is much larger, 55\%, for filter 2899B, used in the 2001 eclipse. A factor of 2.8 better signal is thus expected on this basis alone. We note from Figure~\ref{filter_trans} (upper panel) that a shift in the wavelength of the green line by as much as $\pm 1$~\AA, corresponding to a velocity shift of 57~km~s$^{-1}$, would result in only a very small change of filter transmission; such velocities greatly exceed those seen by the CoMP instrument \cite{tom07}, so any changes of intensity seen in our data are much more likely to result from density changes in the emitting material (the intensity of green-line-emitting material depends on the square of the density).

The manufacturers' specifications for the green-line filters used in the 1999 and 2001 eclipses show a blocking of incident radiation to a very low level, $\approx 10^{-5}$, over the range 2000~\AA\ to 12000~\AA, so there is likely to be negligible response to radiation at, for example, the wavelength of the H$\alpha$ line (6563~\AA). There is no trace of emission from a large prominence, a conspicuous feature during this eclipse, in the green-line images, so there is evidently no response to the continuous emission from the prominence.

\subsection{The Observations}

The instrument, including the components making up SECIS and its optical bench and the heliostat, was transported to the campus of the State University of Zambia some ten days before the eclipse. The roof of the Physics Department of the University of Zambia (latitude $15.4^\circ$ S, longitude $28.3^\circ$ E, altitude 1300~m above sea level) was chosen for the observing site. The orientation of the heliostat and optical bench was first established from calculations using the data of \inlinecite{esp99} for the eclipse at our location, then refined using standard techniques by which star images were tracked with guiding telescopes attached to the heliostat the previous night. Alignment of all optical elements including the heliostat and the exact perpendicularity of the green-line filter was achieved using a small laser mounted at various positions along the optical bench.

The skies were cloudless for the duration of the eclipse totality and for several hours either side of the totality. At the time of mid-eclipse (13:10:56 UT) the Sun's altitude was $31^\circ$ over the north-west horizon. This altitude is less than for our 1999 eclipse measurements ($59^\circ$) but with highly transparent skies this was not a particular problem.

For about an hour before eclipse second contact, the collimator lenses and the lenses in front of each camera were focused with micrometers, the optimum position being obtained by imaging several large sunspots visible on the partially eclipsed Sun images observed through a sheet of thin mylar covering the telescope aperture. Pre-eclipse observations of the full Moon (which has a similar surface brightness as the solar corona) indicated that a neutral density filter with transmission 6.3\% in the path of camera CCD1 would ensure that the white-light images would be correctly exposed. We took precautions to minimize changes in the optics, in particular the focus of the images, to account for the expected drop in temperature during totality, by covering the telescope with aluminium foil and using a high-reflectivity aluminium box containing the cameras. The computer and other electronics were shaded by a wooden box. However, we did not expect the temperature change to be so significant in such a short time that there would be measureable effects on the instrument electronics. Our intention in any case was not to measure absolute flux levels during the eclipse but only the small relative changes possibly arising from coronal heating agents.

Each CCD camera chip has $512 \times 512$ pixels, but the proprietary software operating the cameras limits the chip usable area to $492 \times 504$ pixels. The pixel size corresponds to 4~arcsec $\times 4$~arcsec on the sky, so a region of about 33~arcmin square can be imaged, or approximately one solar diameter. A portion of the visible corona had therefore to be selected. This was undertaken before the eclipse using images from the EIT and MDI instruments on the SOHO spacecraft. These showed that there was a complex of active regions north of Sun centre on 20~June with only minor activity on either the west or east limbs. We decided to point our instrument in the region of the north-east limb near to which was a weak active region (NOAA 9511). In the event, a new and more prominent active region (NOAA 9513) appeared on the north-east limb on 21~June with the bright prominence mentioned above, so our choice of pointing was fortunate.

Our aim was to collect image data over a period a little longer than the predicted totality duration for Lusaka, 194~seconds \cite{esp99}. This could be accomplished by capturing $8000$ images for each channel at a frame rate of 39~frames~s$^{-1}$ -- an odd number of frames~s$^{-1}$ was chosen to avoid the possibility of beating with the frequency of the local electricity supply. With an exposure time  of 22.541~ms for each image and a delay time of 3.1~ms, the duration of the image stream was 205~seconds, or 11~seconds longer than the predicted totality duration. The observations were commenced when the final Baily's bead just south of the Sun's east limb was disappearing. Live images of the Sun during the eclipse totality on the computer screen showed that the pointing of the heliostat remained constant since the last visible image of the partially eclipsed Sun several minutes before eclipse second contact. These images also showed that the cameras performed well and were correctly exposed and in focus. The images were in fact found to have much improved quality over those taken during the 1999 eclipse, with the green-line images having a much stronger signal and with none of the tracking irregularities that occurred during the 1999 eclipse.

\begin{figure}
\centerline{\includegraphics[width=.8\textwidth,clip=,angle=0]{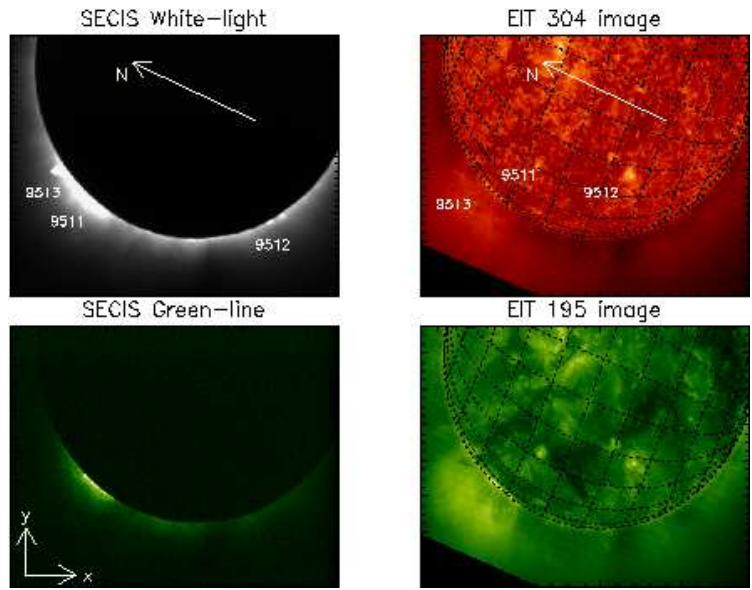}}
\caption{Left 2 panels: Images 1000 from the SECIS white-light channel (upper) and green-line channel (lower), taken $\Delta t = 25.64$~s after the start of the observation sequence).  The orientation is indicated by the arrow, pointing at solar north.  Right 2 panels: SOHO/EIT 304 (upper) and 195 (lower) images, made on 21~June at 01:19:41~UT and 01:13:46~UT respectively. The images have been rotated to the orientation of the SECIS images. NOAA active region numbers are indicated near the corresponding active regions in the EIT image. In the SECIS white-light image, the NOAA active region numbers shown indicate the loops or prominences on the limb that appear to be associated with those active regions.  }\label{SECIS_EIT_images}
\end{figure}

Figure~\ref{SECIS_EIT_images} shows green-line and white-light image number 1000, {\it i.e.} taken $\Delta t = 25.64$~seconds after the start of the image sequence (hereafter we denote times from the image sequence start by $\Delta t$),  in the white-light and green-line cameras. These images showed features on the eastern limb more clearly than in later images when this part of the corona was partly occulted by the Moon. They are compared with SOHO/EIT images in the 304~\AA\ filter (He~{\sc ii} 304~\AA\ Ly$\alpha$ chromospheric emission with some Si~{\sc xi} emission from off-limb coronal structures) and the 195~\AA\ filter (Fe~{\sc xii} 195~\AA\ quiet-coronal emission) taken nearly 12 hours before the eclipse images. The orientation of each image is solar north-west (position angle $\approx 300^\circ$) uppermost. The prominence on the north-east limb  associated with the new active region NOAA 9513 is clearly visible in the white-light images but not in the green-line (Section~\ref{Section_2.2}). EIT images have been rotated to agree with the SECIS image orientation. A number of bright loop structures in the SECIS green-line channel can be identified with active regions evident in the EIT images, marked by their NOAA numbers. The spatial resolution of both white-light and green-line images matches the expected four~arcsec-per-pixel resolution (see Section~3), as is indicated by comparing images at various stages throughout totality. In particular, comparison of the fine coronal structures in images near second and third contacts shows that there was no perceptible change in the image quality, so indicating that the focusing of the lenses did not change appreciably as a result of the temperature drop during totality. The imaged coronal structures are sufficiently fine to show the intensity increases at the numerous places where the loops cross each other, as expected from the optically thin character of the features. The large prominence visible in SECIS white-light images is only faintly visible in the earlier EIT 304 image: it developed into a brighter structure in the intervening few hours.

\begin{figure}
\centerline{\includegraphics[width=0.5\textwidth,clip=,angle=-90]{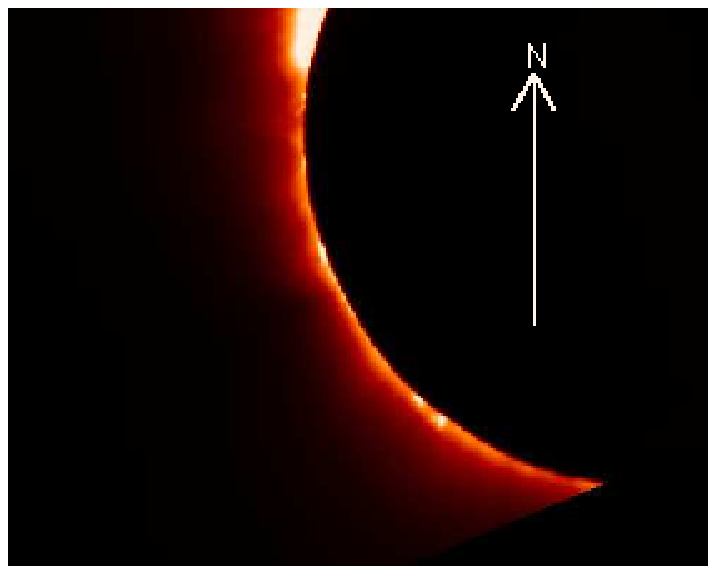}
\includegraphics[width=0.35\textwidth,clip=,angle=-90]{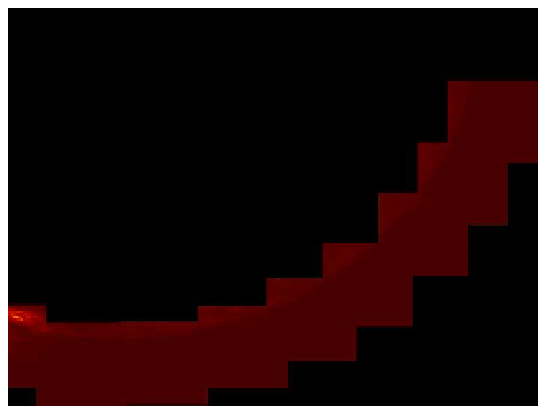}}
\caption{Image 1000 from the SECIS white-light channel (left) and a montage of images around the Sun's west limb made by the SOHO/CDS He~{\sc i} 584~\AA\ window channel taken between 20:03:49~UT on 21~June 2001 and  15:07:53~UT on 22~June 2001 (right). In this figure and the next, solar north is upwards.  }\label{SECIS_CDSHe}
\end{figure}

\begin{figure}
\centerline{\includegraphics[width=0.5\textwidth,clip=,angle=-90]{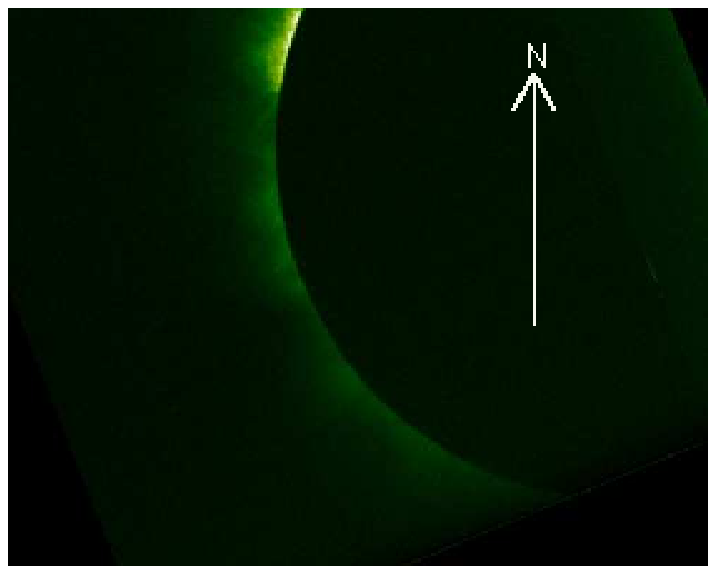}
\includegraphics[width=0.35\textwidth,clip=,angle=-90]{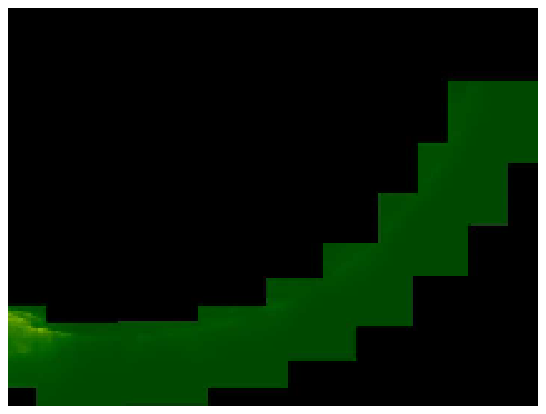}}
\caption{Image 1000 from the SECIS green-line channel (left) and a montage of images around the Sun's west limb made by  the SOHO/CDS Mg~{\sc x} 624~\AA\ channel taken between 20:03:49~UT on 21~June 2001 and  15:07:53~UT on 22~June 2001 (right). }\label{SECIS_CDSMg}
\end{figure}

Observations by teams along the eclipse path were supported by special operations of the SOHO spacecraft; in particular a study was devised in which images in 14 wavelength intervals by the {\it Coronal Diagnostic Spectrometer} (CDS) were made over the entire solar limb, taking over 19 hours starting on 21~June 2001 at 20:03~UT. Figure~\ref{SECIS_CDSHe} shows  SECIS white-light image 1000 showing the loop structures and bright prominence and the corresponding CDS He~{\sc i} 584~\AA\ (peak emitting temperature $\approx 10^4$~K) image. Figure~\ref{SECIS_CDSMg} shows SECIS green-line image 1000 and the corresponding CDS Mg~{\sc x} 624~\AA\ (peak emitting temperature $\approx 1\times 10^6$~K) image.

\section{Stability and Co-alignment of SECIS Images}\label{Section_3}

Visual inspection of the SECIS images in both the white-light and green-line channels showed that the heliostat pointing remained very nearly constant over the period of the observations. To examine the pointing in more detail, computer algorithms were developed that determined the centre of the Moon for each image in the two channels. This was performed with an automated least-squares fitting of a section of a circle to the Moon's limb; to check this, the Markwardt curve-fitting routines available in IDL software were also used. The two approaches led to negligible differences in the results. To refine the alignment still further, to 0.1-pixel accuracy, we performed a two-dimensional cross-correlation analysis of well-defined structures in the corona (the large prominence on the north-west limb was unfortunately too diffuse for this purpose). This resulted in a set of 8000 images in each channel shifted to a common reference system and available for photometric analysis. We were forced to discard the first 1000 images owing to many large vibrations caused by eclipse spectators who were located some distance from our instrument celebrating the start of the totality. We also discarded the last 500 images (because of the bright background formed by the Baily's bead on the solar west limb), leaving a total of 6500 usable images in each channel.

\begin{figure}
\centerline{\hspace*{0.015\textwidth}\includegraphics[width=0.9\textwidth,clip=,angle=-90]{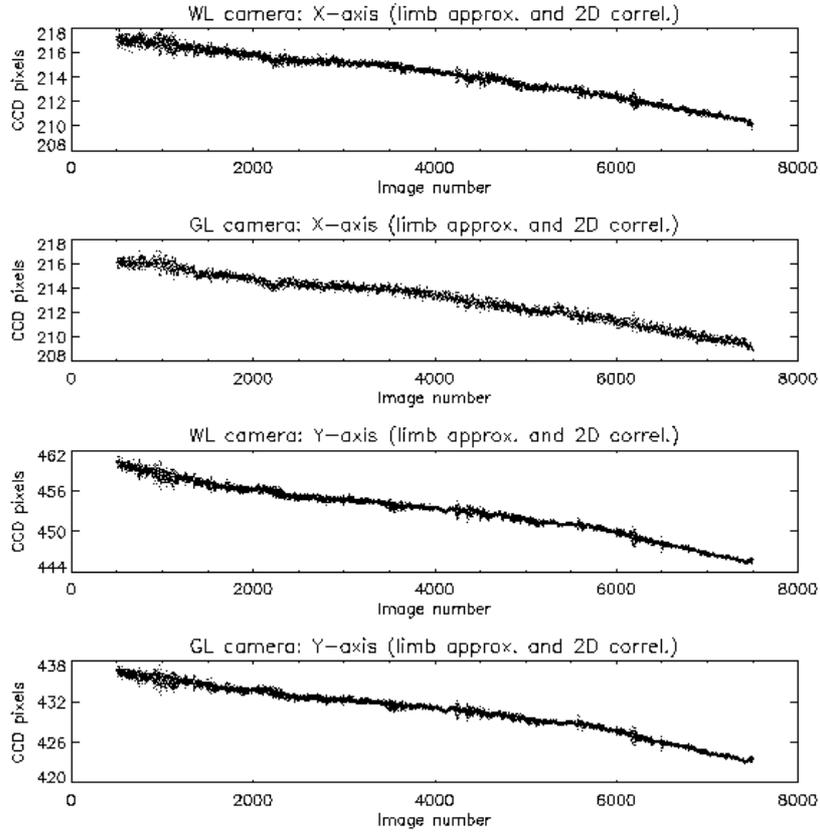}}
\caption{Proper motion of the Moon's centre across CCD2 (green-line or GL camera). Top panel: Motion in $x$. Bottom panel: Motion in $y$. The $x$ and $y$ directions are indicated in Figure~\ref{SECIS_EIT_images}. }\label{motion_of_Moon_centre}
\end{figure}

Figure~\ref{motion_of_Moon_centre} shows the time variations of the Moon's centre coordinates, $x_0$ (upper panel) and $y_0$ (lower panel), for images 500 to 7500 (179.5~seconds) for the green-line camera (CCD2 in Figure~\ref{instrument_layout}). The ($x$, $y$) coordinate system is that given in Figure~\ref{SECIS_EIT_images}. There is a fairly steady but slightly non-linear drift across the field-of-view due primarily to the Moon's motion in the sky (there is also a much smaller motion due to the Sun's motion). Taking the white-light images 2000 to 5000 (a time period of 76.92~seconds), the mean drift is $\Delta x_0 = + 0.029$~px s$^{-1}$ and $\Delta y_0 = - 0.053$~px s$^{-1}$. The steady drift of the Moon's centre is a marked improvement over the irregular motions present in our 1999 eclipse data, evidently as a result of the metal box containing the heliostat drive motor and electronics. Our analysis indicates that the Moon's radius is $r = 241.9 \pm 0.3$~px. Using the data of \inlinecite{esp99}, the angular radius of the Moon at mid-eclipse from Lusaka was 987~arcseconds, so the scale of our images is 4.07~arcseconds per pixel, in almost exact agreement with the estimates ($4.09 \pm 0.02$~arcseconds per pixel) from our earlier work \cite{rud04}.  In addition to the Moon's proper motion, there are small (about $\pm 0.3$~px), short-term (up to ten data points or 0.25~seconds) excursions in each direction which are probably due to the small errors in the heliostat tracking.

An analysis of the changes in the Moon's centre $x$- and $y$-coordinates was undertaken to search for periodic or erratic image motions that could affect the photometric analysis. Such periodicities might, for example, have been caused by constant-frequency vibrations in the building structure on top of which the equipment was mounted, or by some instabilities of the heliostat's stepper-motor driver or gear. In this analysis and the analyses of the photometric data (Section 4), we made use of wavelet transforms to search for oscillations in the data varying over time as well as standard Fourier transforms. This technique has become standard in many fields including solar physics: see {\it e.g.} \inlinecite{tor98}; \inlinecite{boc95}; \inlinecite{sta97} for further details. The data in our case consist of measured quantities $S_n$ at times $t_n$ separated by a constant time interval 25.64~ms. With wavelet transforms, information can be obtained about temporal variations of the observed signal with a time or frequency analysis by computing the correlation of the signal and a chosen wavelet function $\psi$ that is localized in time and frequency. The wavelet transform is then defined by a convolution of the data with a scaled version of the wavelet function $\psi [(t - t_0)/s]$; by varying the wavelet scale $s$ (related to period or the inverse of frequency), a wavelet power spectrum can be constructed showing the power of any features present as a function of $s$ and of time $t$. In the code we wrote, IDL routines were used with a Morlet wavelet (see \opencite{tor98}), following extensive use in solar problems, together with confidence limits assigned to features in the wavelet power spectra that appeared to be significant. Also, cones of influence were defined outside of which edge effects are important.

\begin{figure}
\centerline{\hspace*{0.015\textwidth}\includegraphics[width=0.7\textwidth,angle=0,clip=]{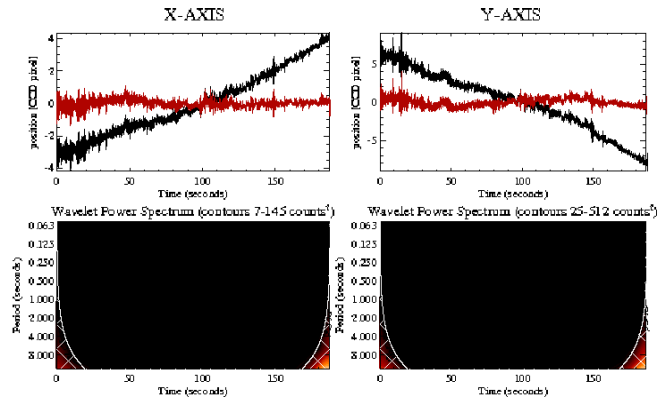}}
\caption{Wavelet analysis of the Moon's proper motion, shown in Figure~\ref{motion_of_Moon_centre}, in the $x$ (left) and $y$ directions.  Top 2 panels: Black curves show motion about the mean $x$ and $y$ positions; green curves show motion with main linear trend removed. Bottom 2 panels: Wavelet power spectra corresponding to the upper panels showing no apparent power except at very low frequencies (light areas outside the cone of influence, shown as the cross-hatched areas). Global power spectra (power summed over the entire span as a function of period) are shown to the right, with 95\% significance level. }\label{wavelet_Moons_motion}
\end{figure}

The results of the analysis for the measured positions of the Moon's centre are shown in Figure~\ref{wavelet_Moons_motion}, where the $x$ and $y$ motions are plotted with time in the upper panels and the wavelet analysis in the bottom panels. There is no indication of any periodicity in the wavelet power spectra in the frequency range 0.1--14~Hz.

We neglected the extremely small rotation of the images caused by the heliostat over the totality period.

\section{Photometric data analysis}\label{Section 4}

\subsection{Green-Line Image Photometry and Noise Assessment}

\begin{figure}
\centerline{\hspace*{0.015\textwidth}\includegraphics[width=0.8\textwidth,clip=,angle=0]{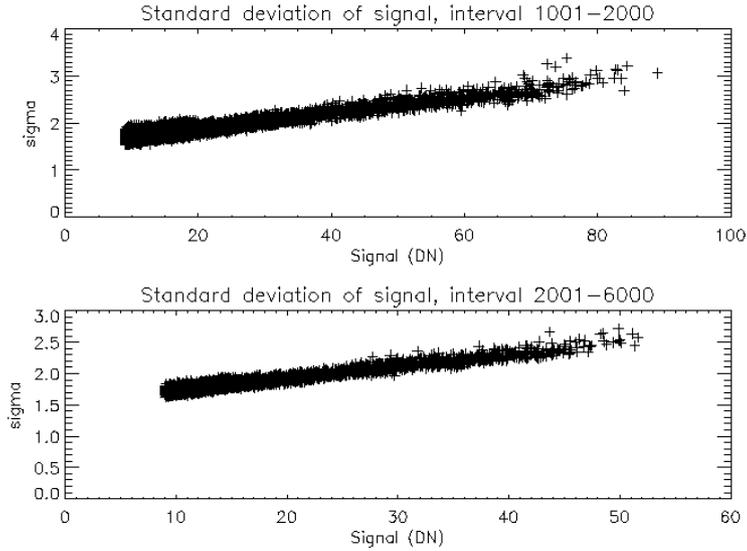}}
\caption{Standard deviation in the signal for all pixels having strength $\geqslant 10$~DN. Top panel: Images 1001--2000. Bottom panel: Images 2001--6000.  }\label{signal_noise}
\end{figure}

To search for periodicities in the intensity of the green-line images, we must assess the likely level of instrumental or intrinsic noise. There are three likely sources: $i$) camera dark current noise; $ii$) camera read-out noise; $iii$) ``shot" noise, {\it i.e.} the inherent noise of the source arising from the statistical nature of arrival of photons from the Sun. The output of each of the cameras is nominally a 12-bit digital number (DN) for each pixel, but the two least significant bits are due to noise; the data are therefore effectively digitized to ten bits, or 1024 levels of signal. The standard deviation of the signal was measured for each pixel having a signal strength of $\geqslant 10$~DN, well above the dark-current level ($\approx 2$~DN: Section~4.2), so is likely to be mostly solar in origin. This was performed for two time periods, images 1001--2000 ($\Delta t = 25.6$--51.3~seconds) and images 2001--6000 ($\Delta t = 51.3$--153.8~seconds), removing pixels that were occulted by the Moon during each period. During the first period, the bright active region complex NOAA 9511/9513 was visible, with pixels having $\gtrsim 80$~DN. During the second period, this bright region was occulted by the Moon so that pixels only up to $\approx 50$~DN were visible. The standard deviation [$\sigma$] for each pixel is plotted against the signal strength (DN) in Figure~\ref{signal_noise} for the two time intervals.

The increase of $\sigma$ with signal strength in both time intervals (see Figure~\ref{signal_noise}) is due to a combination of shot noise, which increases as the square root of the intrinsic signal, and instrumental noise made up of dark current and read-out noise which may be considered approximately constant. With the relation shown in Figure~\ref{signal_noise} assumed to be linear, we find that the instrumental noise is 1.4~DN. Defining the signal-to-noise ratio (SNR) in the usual way to be the ratio of the mean pixel DN to the standard deviation [$\sigma$] in a given interval, we found that the SNR is 29 for pixels in the active regions NOAA 9511/9513, but correspondingly lower for pixels near the Moon's limb.

Some points in the plots, particularly the upper plot for the earlier time interval, are well above the linear relation. We identified these pixels and investigated any periodic trend in their signal (see below, Section~\ref{first_anal}).

\subsection{Image Flat-Fielding and Dark-Current Subtraction}

Dark-current and flat-field data were collected by taking several series of 500 dark-current and 100 flat-field exposures, each exposure time being 22.5~ms. Through some confusion in the storing of the eclipse data after totality, the dark-current exposures were performed later than was expected, near the time of local sunset, when the cameras were cooler than at the time of the eclipse. There was insufficient time to take flat-field exposures until the following morning; these were accomplished by offsetting the heliostat to featureless blue sky well away from the Sun. A repeat of dark-current exposures was performed at a temperature similar to that during the previous day's eclipse.

As mentioned in Section~2.3, the total area of each camera chip nominally available for analysis is $492 \times 504$ pixels. However, parts of the image area of each camera are affected by a non-uniform response of the camera chip. For camera CCD1 (white-light channel), the ten left-most columns are strongly non-uniform, while for camera CCD2 (green-line), the 15 left-most columns and a wide band consisting of the uppermost 112 rows are non-uniform and are affected by a imprinted fixed pattern. As in previous runs of the cameras (including the 1999 eclipse), these regions were excluded from the analysis, leaving a usable area measuring $476 \times 376$ pixels. To minimize the effect of this reduced area of the camera chips, we used previous experience from the 1999 eclipse to orient both cameras such that the Moon's disk at mid-eclipse would cover the unusable area while the corona would cover the usable areas of the CCDs.

\begin{figure}
\centerline{\hspace*{0.015\textwidth}\includegraphics[height=0.5\textwidth,width=0.5\textwidth,clip=,angle=-90]{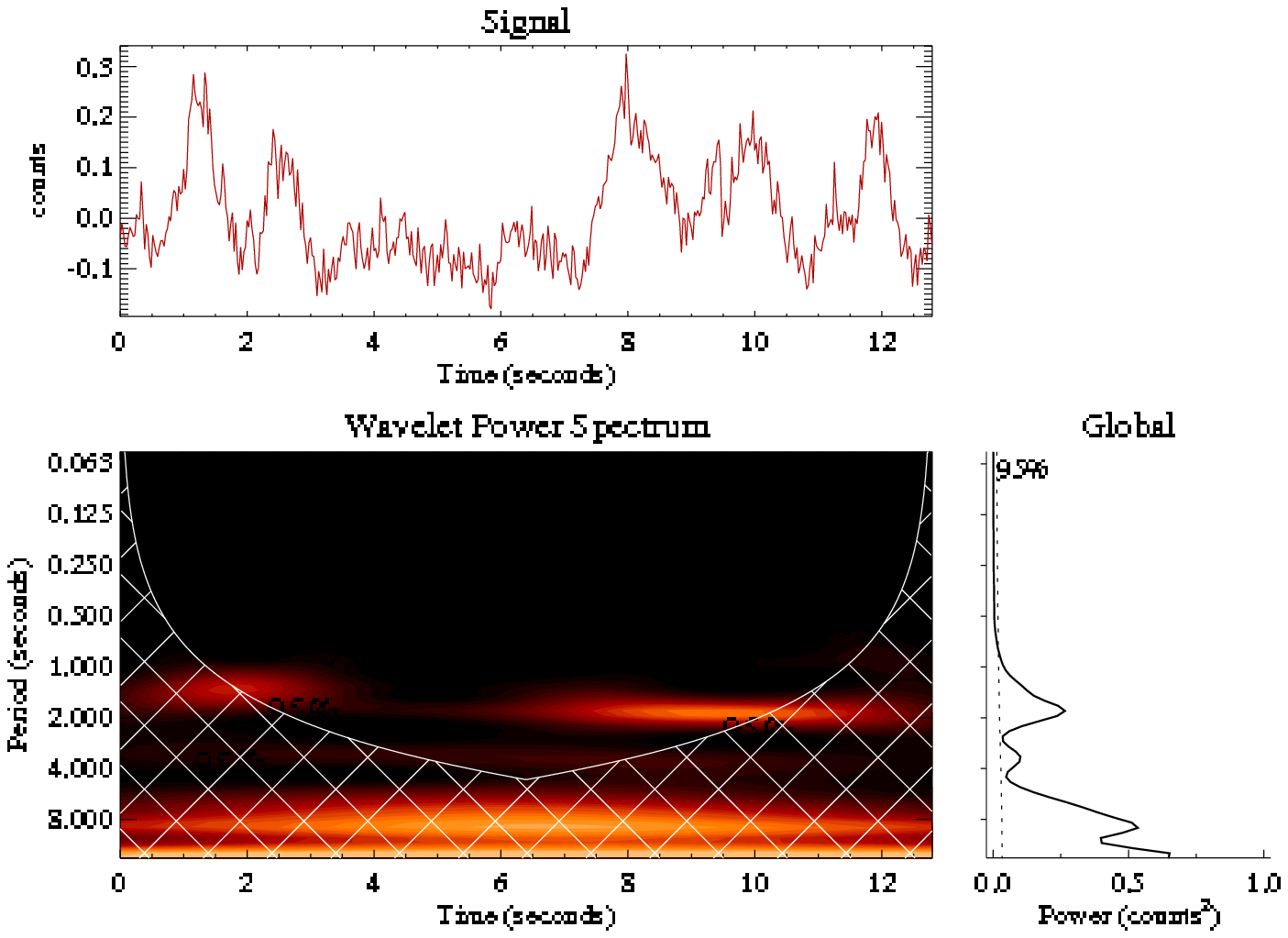}
\includegraphics[height=0.5\textwidth,width=0.5\textwidth,clip=,angle=-90]{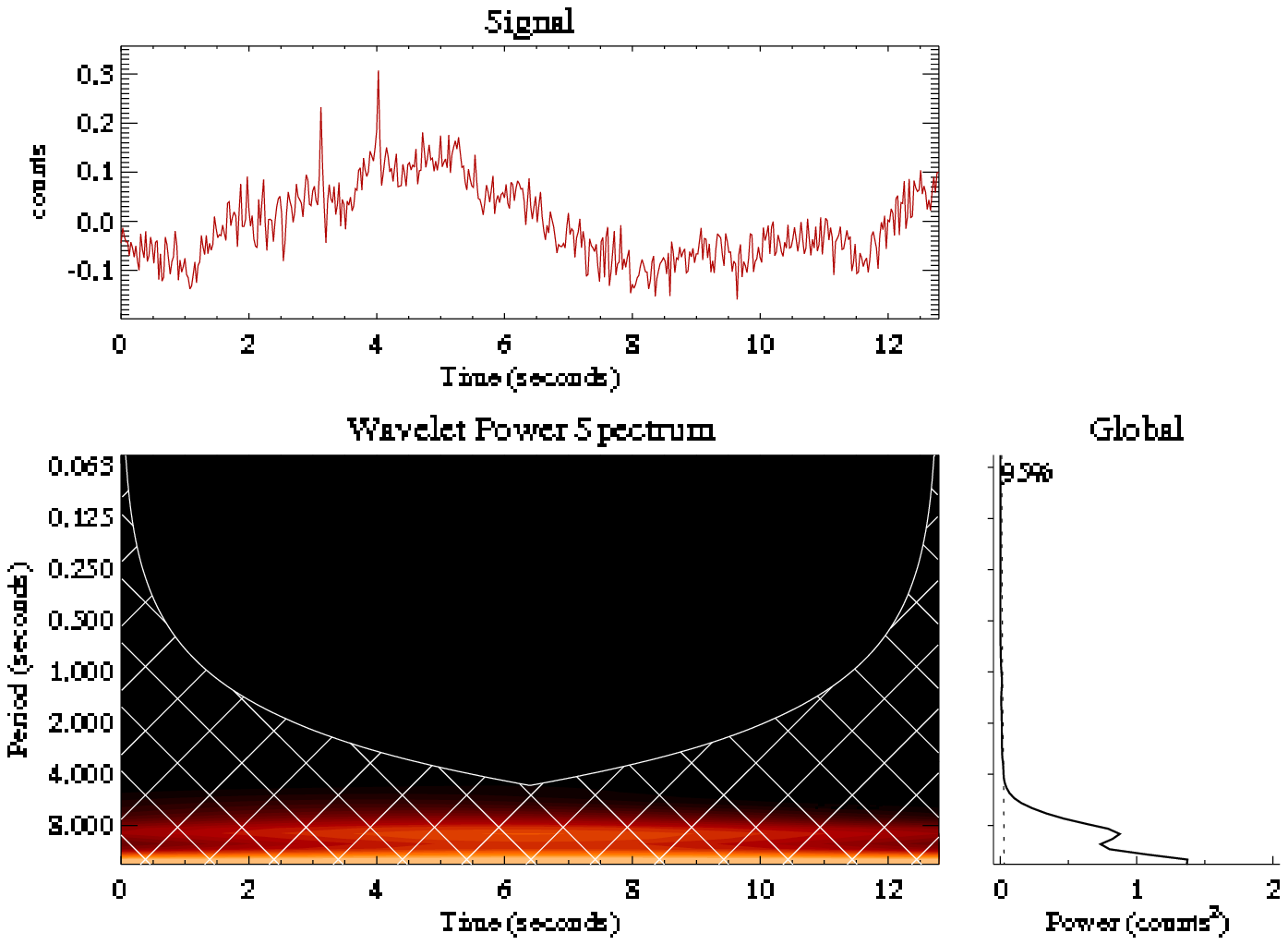}}
\caption{Wavelet power spectra of dark-current exposures taken in the green-line channel (left lower panel) and white-light channel (right lower panel). The analysis is based on 500 exposures with combined exposure and delay times of 25.64~ms, or a total time span of 12.8~seconds. Enhanced power is shown by light areas, with time (in seconds) during the exposures in the abscissa direction and period (in seconds, {\it i.e.} the inverse of frequency in Hz) in the ordinate direction. The upper plots are the signal light curve with mean level subtracted and a bell-shaped function applied to suppress jumps at the start and end of each time span. Global spectra are indicated to the right of the lower plots. }\label{wavelet_DC}
\end{figure}

Analysis of these exposures revealed some gain instabilities having their origins in the electronics. The mean dark-current and flat-field signals within the usable areas of each camera were calculated for each image, and a wavelet analysis of the time series
was undertaken. This analysis showed some periodic tendencies. Figure~\ref{wavelet_DC} shows an example of a wavelet power spectrum of this analysis for the dark current images. There is a suggestion of periods between one and two~seconds  for the green-line channel, significant at the 95\% level. There is also a longer-period oscillation with period $\approx 8$~seconds (0.13~Hz) in both channels, but which is outside the wavelet ``cone of influence", {\it i.e.} where periodicities do not have statistical significance because they are substantial fractions of the time span of the exposures (12.8~seconds).  The one-to-two~second periodicity is clearly of importance in analyzing green-line data for regions of low signal strength. The average photon count rate in the dark-current exposures is about 2~DN, so the investigation of periodicities in the eclipse data is only possible for regions with count rates of much more than 2~DN.

We averaged each flat-field and dark-current sequence to form a single flat-field and dark-current image which were then applied in a standard way to the photometric data.

\subsection{Selection and Analysis of Data}

The final data set used to search for periodicities in the coronal brightness thus consisted of images 1000 to 7500 ($\Delta t = 25.64 - 192.31$~seconds), all spatially shifted to a common reference system and corrected for flat fielding and dark current.  Green-line data were analyzed in two stages, one involving a selection of the available data, the other a more comprehensive analysis of all pixels having significant signal level within a broad area above the Moon's limb. The white-light data were also analyzed as a check on results obtained from the green-line analysis.

\subsection{Analysis of Selected Locations}\label{first_anal}

We took as a starting point in the analysis of SECIS data the results from the CoMP instrument of \inlinecite{tom07}, in which wave-like motions were observed proceeding along magnetic flux lines near active region loops. These had much lower frequencies than those for which we were searching for in our data, but nevertheless provided a guide to where we should focus our attention.

\begin{figure}
\centerline{\hspace*{0.015\textwidth}\includegraphics[width=0.6\textwidth,clip=,angle=-90]{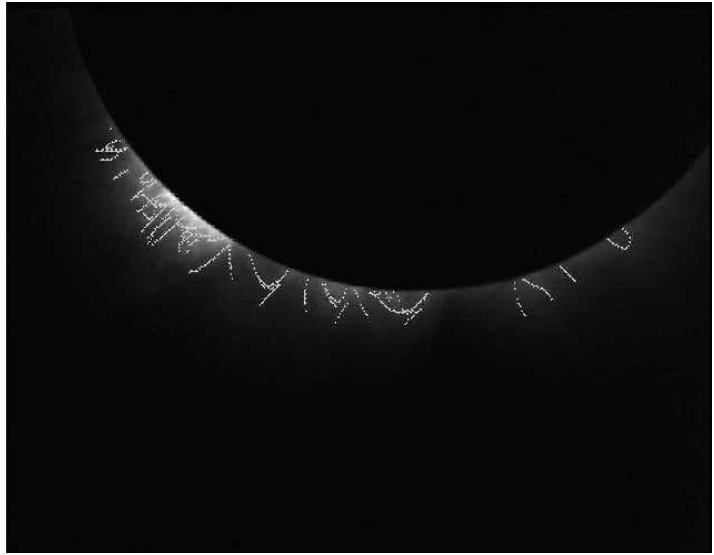}}
\caption{Locations of the 1385 pixels selected for the first analysis. These are indicated in the green-line image 1000 as white dots, and lie in and around loop-like structures off the limb.  }\label{1385_pixels}
\end{figure}

We selected single pixels for analysis as follows:  {\it i}) pixels at specific locations in and around the loops making up the active region complex on the north-west limb (NOAA 9511/9513); {\it ii}) pixels within loop structures on the south-west limb with no strong association with active regions; {\it iii}) signal strength $\geqslant 9$~DN. A total of 1385 pixels was selected. Figure~\ref{1385_pixels} shows the locations of these pixels in green-line image 1000. Wavelet power spectra and plots of Fourier power for image set 1001--2000 ($\Delta t = 25.64 - 51.28$~seconds), generated by the automated routine, were then scanned by eye for any obvious examples of periodicities, particularly in the wavelet  power spectra where such periodicities are manifested by horizontal bands of enhanced power at specific frequencies persisting for a significant fraction of the  time interval examined.

\begin{figure}
\centerline{\hspace*{0.015\textwidth}\includegraphics[width=0.3\textwidth,clip=,angle=-90]{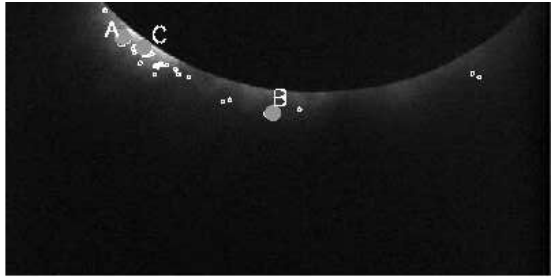}}
\caption{Locations of pixels in SECIS green-line images 1001--2000 that show enhanced Fourier power: symbols are small open circles (for initial selection of 54 pixels with enhanced power in waveletgrams, selected by eye) and large full circles for pixels marked A, B, and C (see main text for further details).  }\label{54_selected_pixels}
\end{figure}

\begin{figure}
\centerline{\hspace*{0.015\textwidth}\includegraphics[width=0.5\textwidth,clip=,angle=-90]{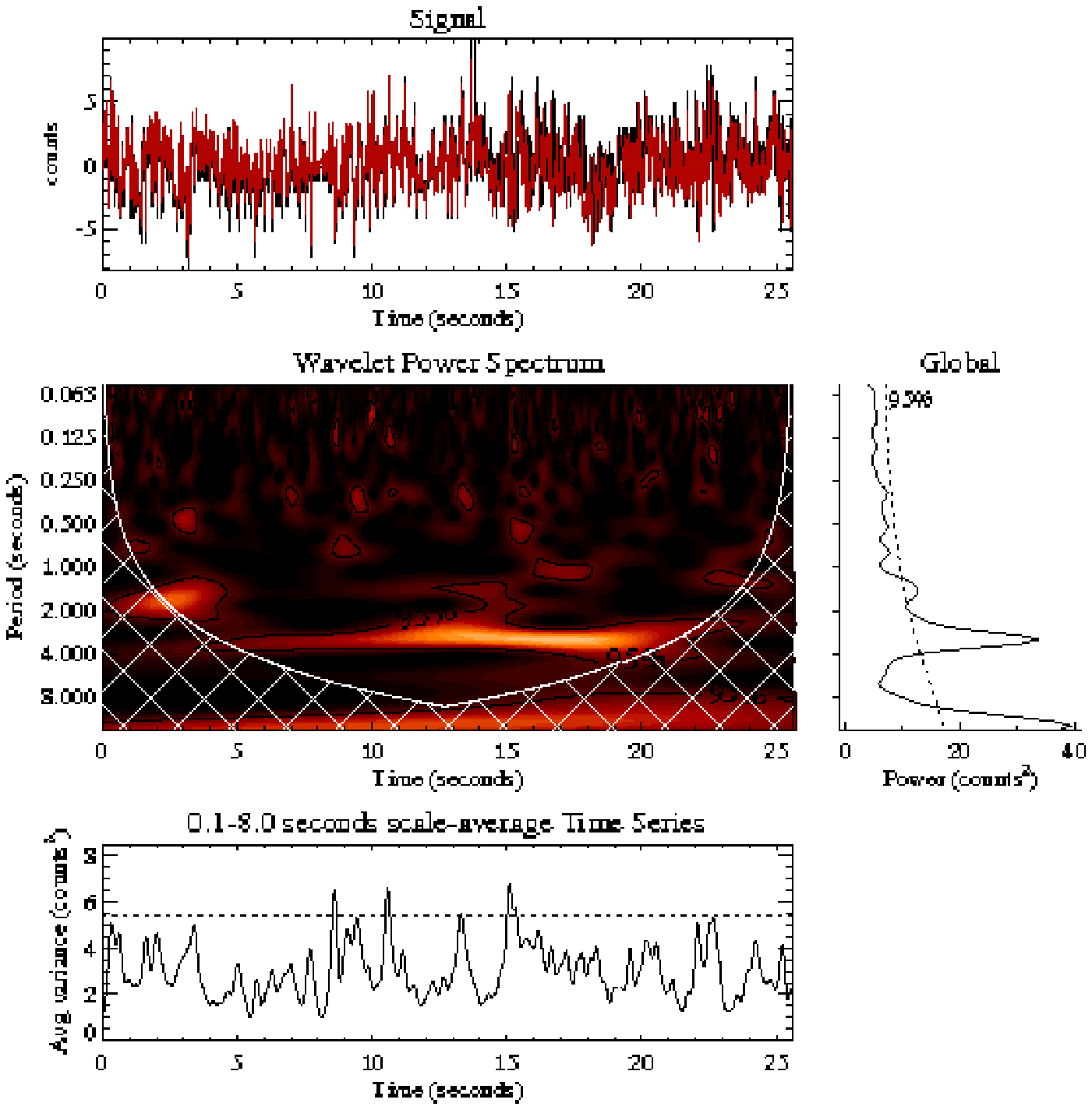}
\includegraphics[width=0.5\textwidth,clip=,angle=-90]{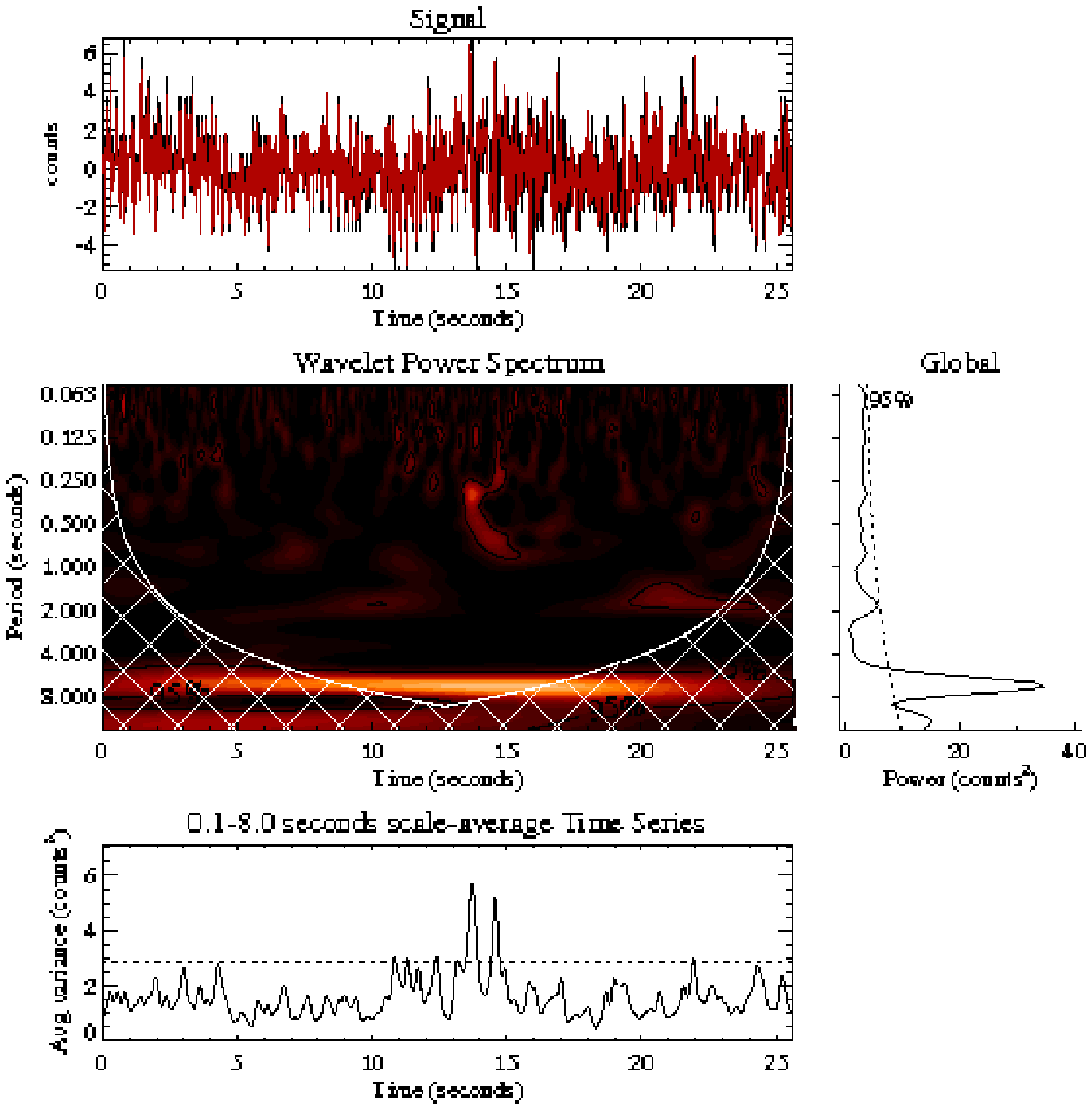} }
\caption{Wavelet power spectra for pixels A (left panels) and B (right panels), images 1001--2000, showing enhanced power at 0.36~Hz and 0.17~Hz respectively (see Figure~\ref{54_selected_pixels} for locations of these pixels). The panels of the wavelet power spectra are arranged as in Figure~\ref{wavelet_DC}. The SFT power spectrum has peaks $> 3 \sigma$ significance for each of the enhanced frequencies. }\label{waveletgrams_2 locs_1001_2000}
\end{figure}

Some 54 pixels with possible periodicity were evident from this first examination, the locations of which are shown in Figure~\ref{54_selected_pixels}. The wavelet power spectra of two particularly striking examples are illustrated in Figure~\ref{waveletgrams_2 locs_1001_2000} (the plots are arranged as in Figure~\ref{wavelet_DC}). The locations of these are shown as large filled circles marked A and B in Figure~\ref{54_selected_pixels}, and are in contrasting locations. There is a strong suggestion of periodicity for these two pixels; the calculated standard Fourier transform shows power exceeding three standard deviations ($3 \sigma$, {\it i.e.} having $> 95$\% significance for a normal distribution).  For pixel~A, the standard Fourier transform (SFT) peaks at a period equal to 2.78~seconds (0.36~Hz), with mean signal strength 42~DN, considerably above the dark-current level of 2~DN. Pixel~A is located on the north side of NOAA 9511/9513, within what SECIS images show to be a complex loop structure. The SOHO/EIT and CDS images (Figures~\ref{SECIS_EIT_images}, \ref{SECIS_CDSMg}), although not simultaneous with the SECIS images, confirm that there are many loops within this active region complex. For pixel~B, the SFT peaks at a period equal to 5.9~seconds (0.17~Hz), with mean signal strength 24~DN. This pixel is located in a more isolated region, far from the nearest active region loops. The frequency in this case is rather near the boundary of the cone of influence. In addition, the signal is weaker than for pixel~A, but it is sufficiently strong that we are confident that the observed periodicity is not attributable to any observed in the dark current exposures with period 8.3~seconds (0.12~Hz) (see Figure~\ref{wavelet_DC}).

\begin{figure}
\centerline{\hspace*{0.015\textwidth}\includegraphics[width=0.5\textwidth,clip=,angle=-90]{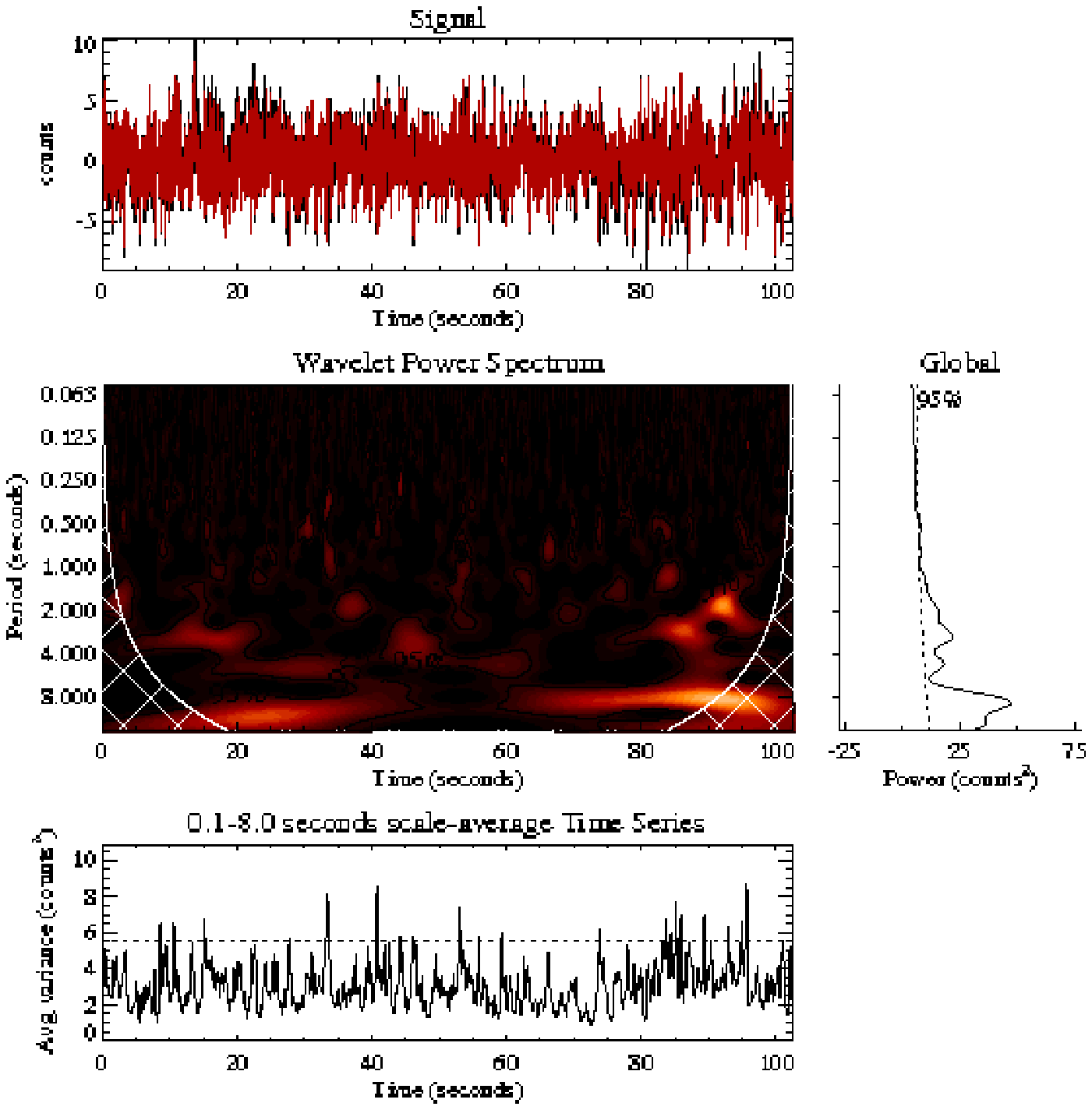}
\includegraphics[width=0.5\textwidth,clip=,angle=-90]{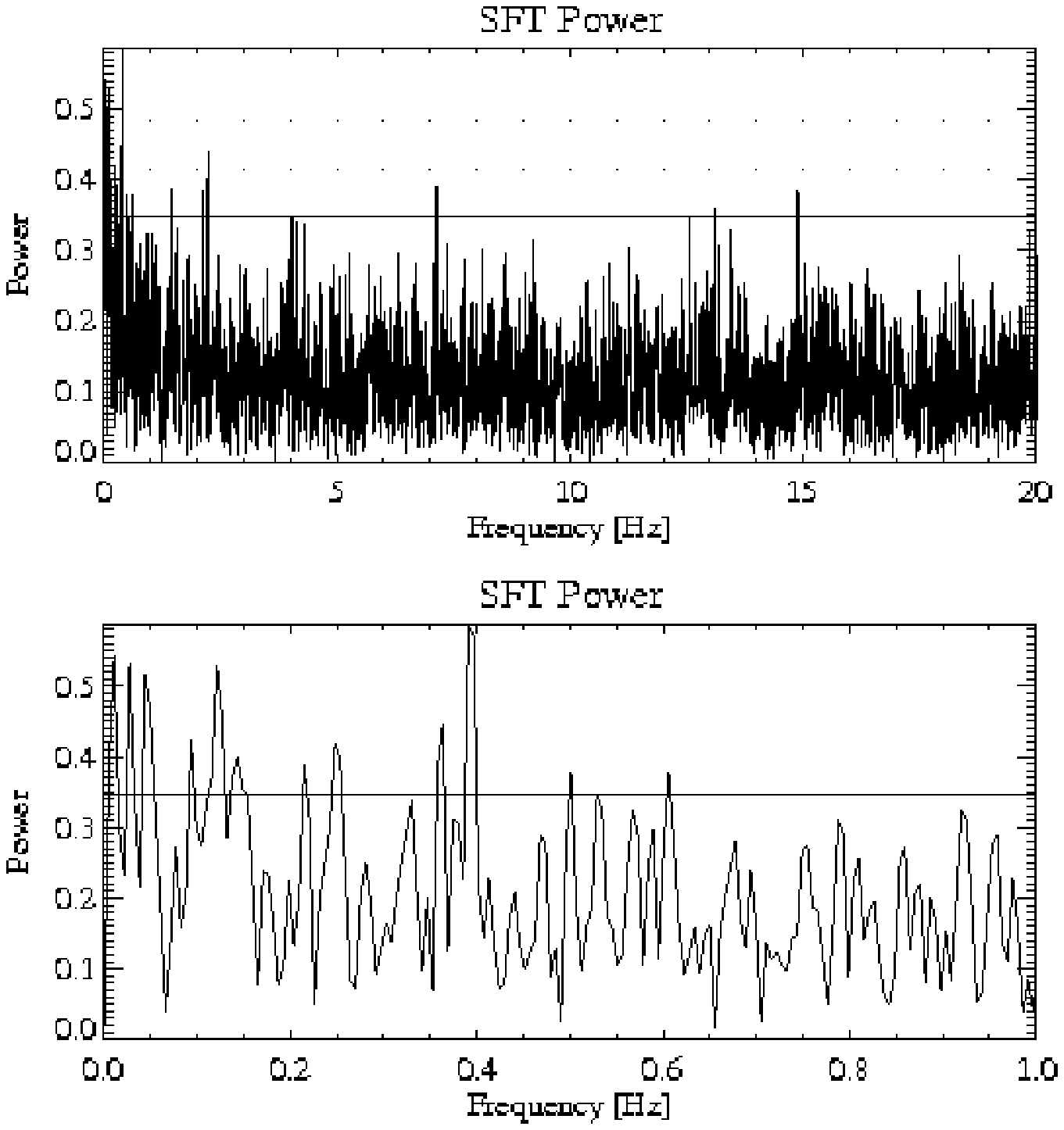} }
\caption{Wavelet power spectrum for  the pixel A for images 1001--5000 (left) and standard Fourier transform (right). The strong three-second periodicity in Figure~\ref{waveletgrams_2 locs_1001_2000} is not evident in later time spans (2001--3000, 3001--4000, and 4001--5000), and has no significance over the 1001--5000 range. The SFT peak at the corresponding 0.33~Hz frequency has $<3\sigma$ significance.  }\label{waveletgram_A}
\end{figure}

\begin{figure}
\centerline{\hspace*{0.015\textwidth}\includegraphics[width=0.5\textwidth,clip=,angle=-90]{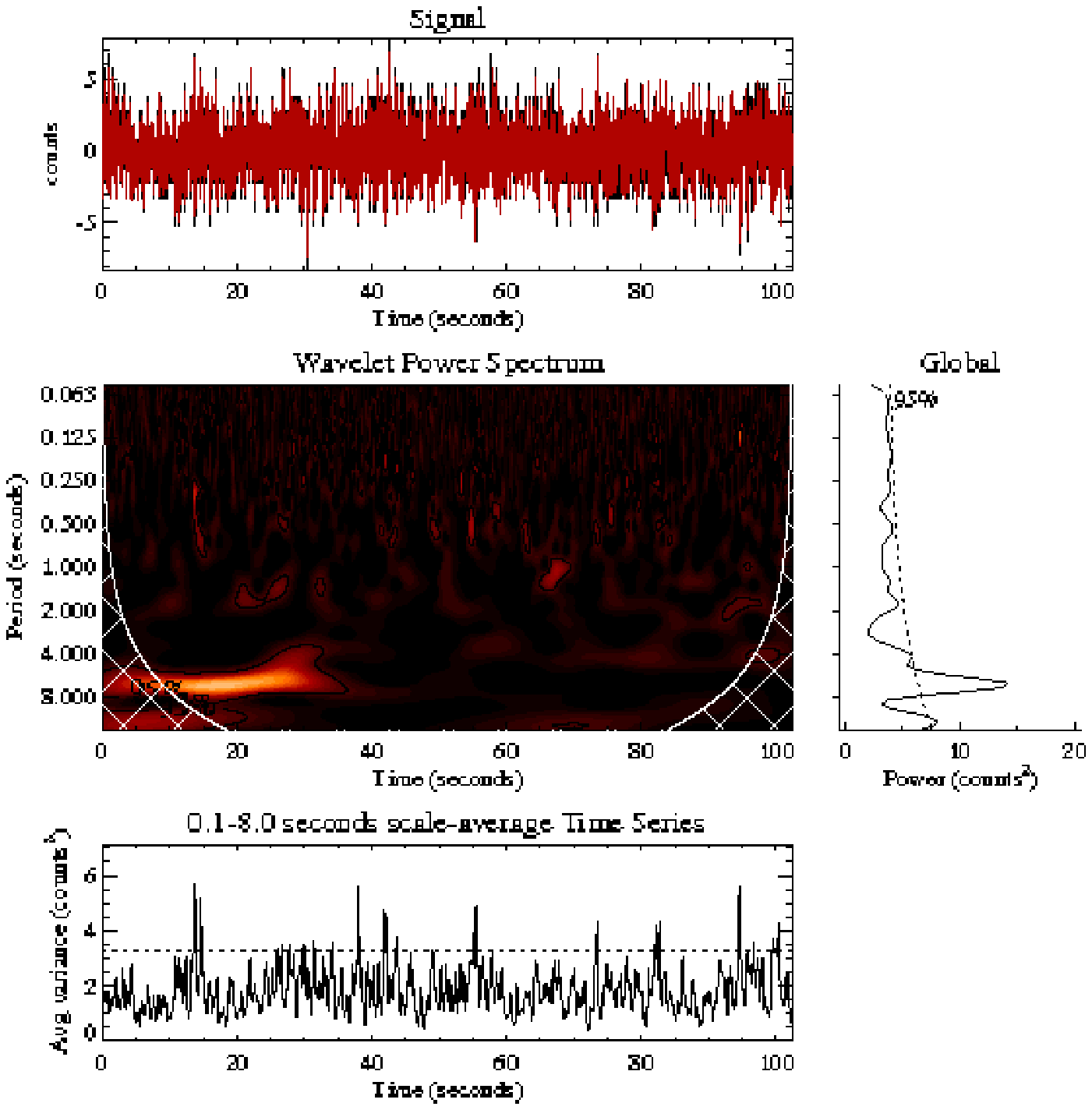}
\includegraphics[width=0.5\textwidth,clip=,angle=-90]{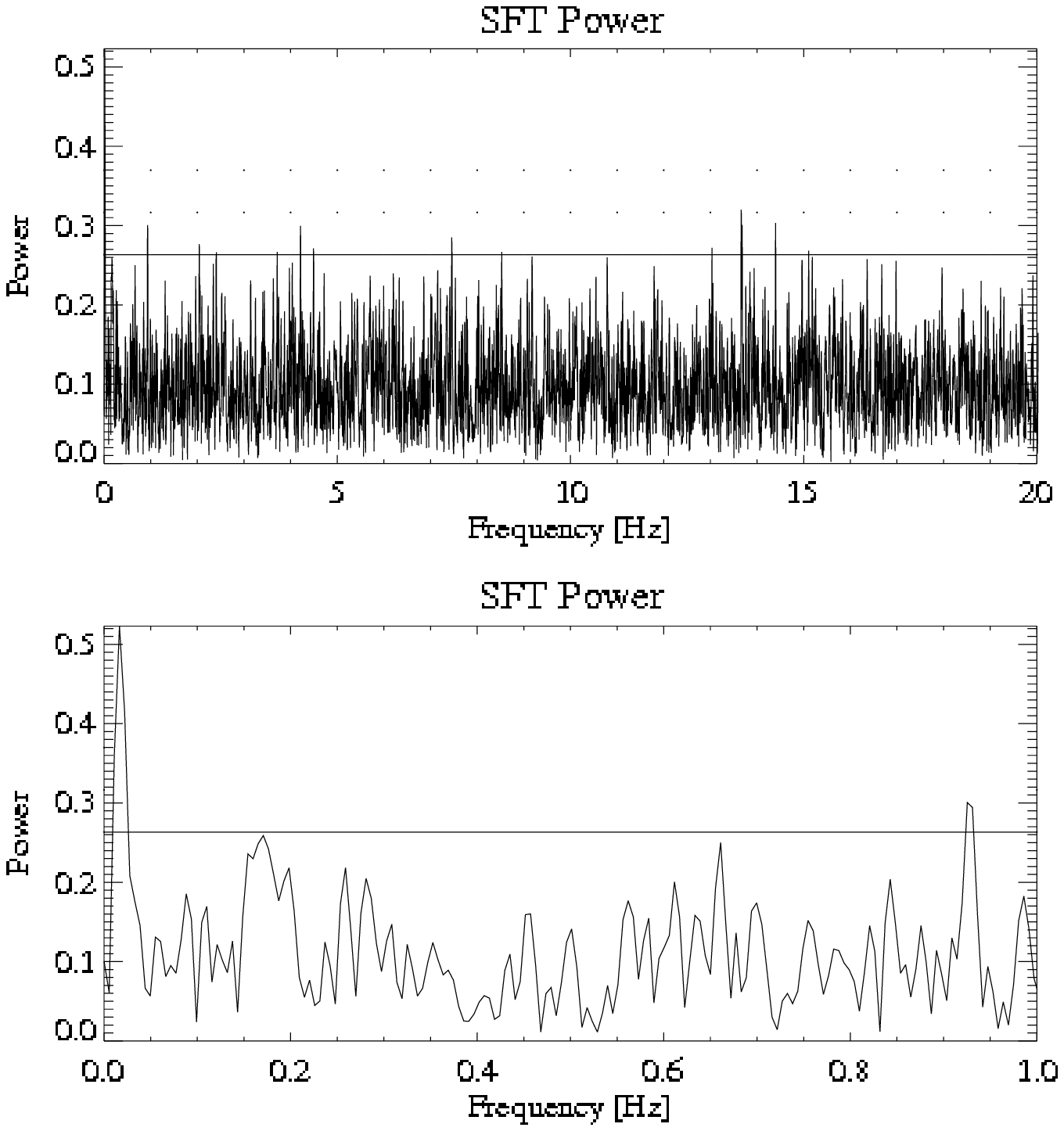} }
\caption{Wavelet power spectrum  for pixel B for images 1001--5000 (left) and standard Fourier transform over frequency ranges 0--20~Hz and 0--1~Hz (right panels).  The strong periodicity indicated in Figure~\ref{waveletgrams_2 locs_1001_2000} is much less evident in the wavelet power spectrum and the corresponding SFT peak at 0.17~Hz (period 5.9~seconds) is slightly less than $3\sigma$ significance.  }\label{waveletgram_B}
\end{figure}

Any persistence in the periodicities shown in the 54 pixels taken from this first analysis was investigated using longer time spans. It was found that cases showing marked evidence for periodicities in images 1001--2000 showed much less evidence over longer time spans. As an example, we consider the two pixels in Figure~\ref{waveletgrams_2 locs_1001_2000}. The wavelet power spectra for images 1001--5000 ($\Delta t = 25.64 - 128.21$~seconds) for pixel~A, together with standard and fast Fourier transforms (SFT, FFT), are shown in Figure~\ref{waveletgram_A}, and for pixel~B in Figure~\ref{waveletgram_B} (later time spans were not possible because the Moon had occulted this region). The six-second period that was so strong in Figure~\ref{waveletgrams_2 locs_1001_2000} is now much less significant, with the corresponding SFT peak at 0.17~Hz having $< 3 \sigma$ significance. Likewise, the strong three-second period in the pixel~A in Figure~\ref{waveletgrams_2 locs_1001_2000} is no longer evident in later time spans; there is no significant peak in the SFT periodogram at the corresponding frequency 0.3~Hz. Neighbouring pixels were also examined over various time spans but they also do not show any strong periods, in particular those with periods indicated in Figure~\ref{waveletgrams_2 locs_1001_2000}.

\begin{figure}
\centerline{\hspace*{0.015\textwidth}\includegraphics[width=0.5\textwidth,clip=,angle=-90]{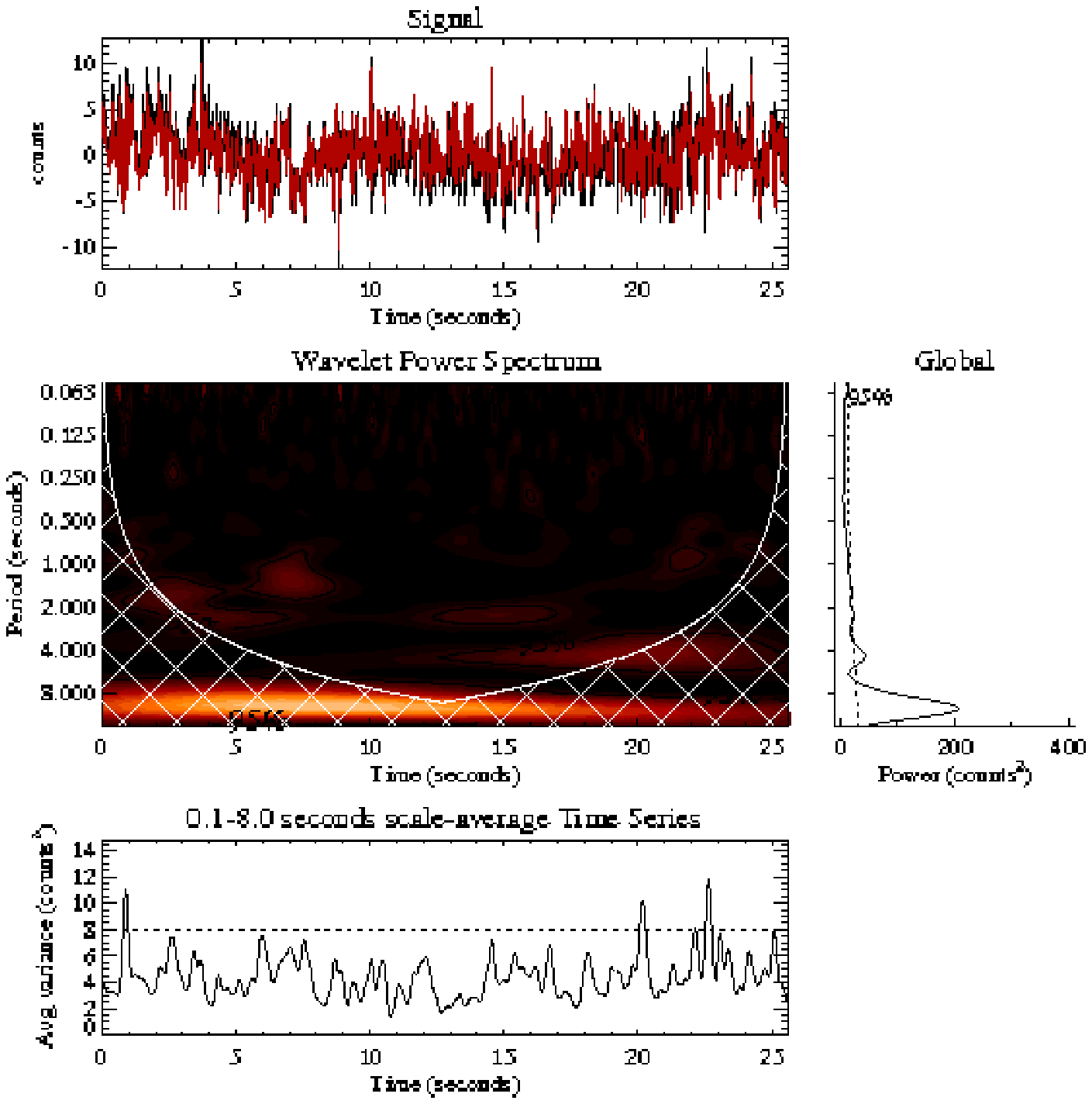}
\includegraphics[width=0.5\textwidth,clip=,angle=-90]{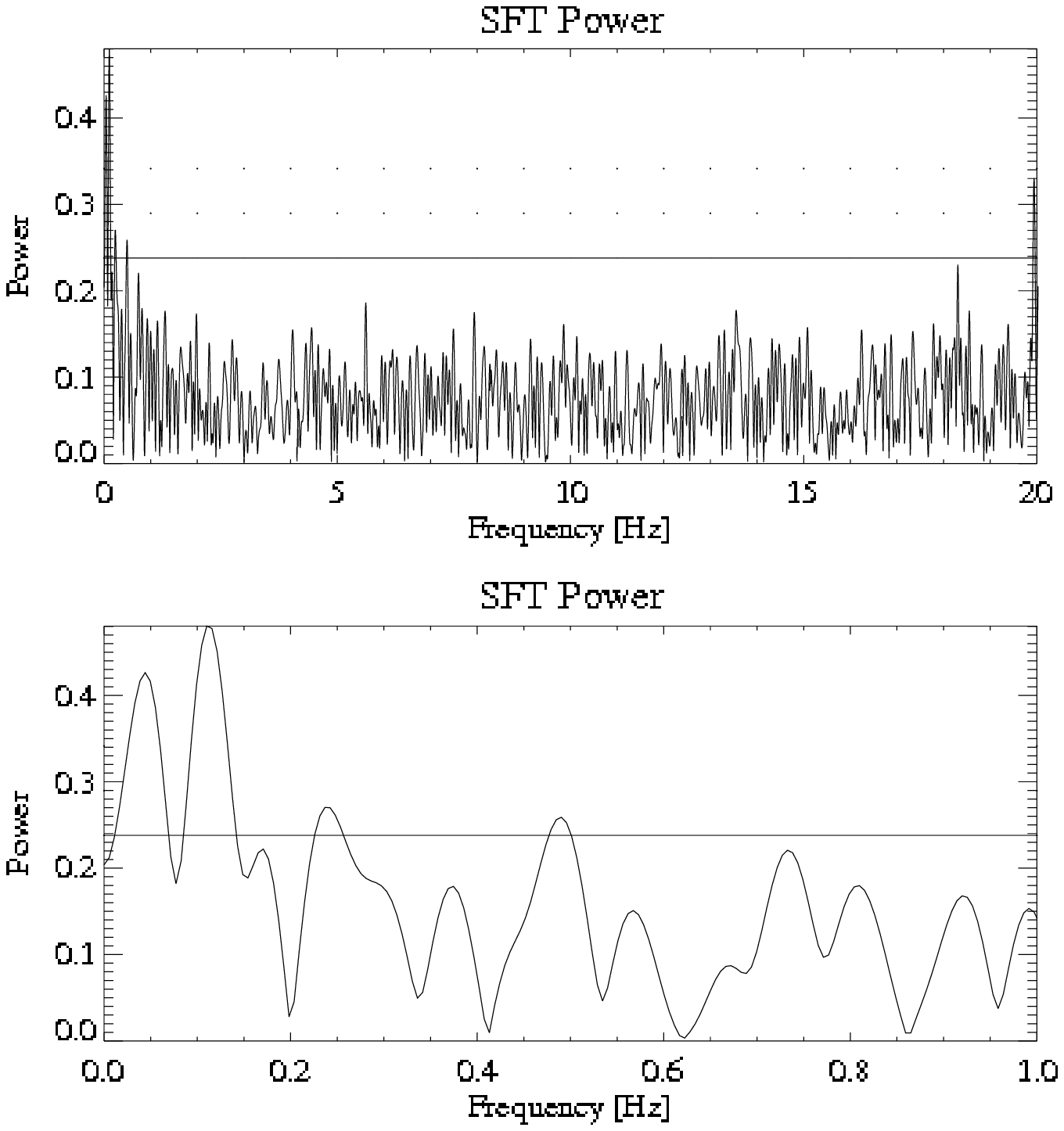} }
\caption{Wavelet power spectrum  for pixel C for images 1001--2000 (left) and standard Fourier transform over frequency ranges 0--20~Hz and 0--1~Hz (right panels). This pixel was occulted soon after image 2000 so the strong periodicity indicated here for a period of $\approx$ nine~s cannot be corroborated in later data. }\label{waveletgram_C}
\end{figure}

Another example of a pixel with possible periodicity is the one marked C in the plot of $\sigma$ against signal strength in Figure~\ref{signal_noise} and in Figure~\ref{1385_pixels}. The wavelet spectrum and Fourier analysis for images 1001--2000 are shown in Figure~\ref{waveletgram_C}. The value of $\sigma$ is well above the linear relation suggested in Figure~\ref{signal_noise}, and the wavelet analysis suggests a periodicity of $\approx 10$~seconds. As this period is a large proportion of the total time span of the observations (25.6~seconds), the periodicity is outside the cone of influence in the wavelet power spectrum. Up to five neighbouring pixels also show similar periodicity. However, as all of these pixels are occulted by the Moon soon after this interval, we cannot confirm the reality of this periodicity and we are forced to suspend judgement on their possible solar origin.

The conclusion from this first analysis is that periodicities identified by eye in the 1385 pixels located at likely positions in and around active regions and more isolated loops are chance occurrences, with none of them persisting for longer than about 20 to 30 seconds.

\subsection{Comprehensive Analysis}\label{comp_anal}

For three image time spans of various lengths (images 1001--6000, time span of 128.2~seconds; 1001--4000, time span of 76.9~seconds; 1001--2000, time span of 25.6~seconds), we selected  crescent-shaped areas around the Moon's limb out to a radial distance of 30~pixels (approximately 120~arcsecs) and with the inner boundary defined such that no pixels were occulted by the Moon over the time series. We then analyzed the wavelet power spectra for all pixels within this area having signal strength $\geqslant 10$~DN.  Most of the analyzed pixels have signal strengths between about 16~DN and 30~DN, with many having signal strengths as large as $\approx 60$~DN. The total number of pixels within the crescent-shaped area for images 1001--2000  is $29\,518$, with $10\,714$ having significant ($\geqslant 10$~DN) signal.


A more comprehensive analysis (Section~\ref{comp_anal}) of all pixels having a signal of $\geqslant 10$~DN (dark current level is $\approx$ two~DN)  similarly does not show up any significant periodic tendency that is sustained for more than a few seconds. There are, however, many pixels having periodic tendencies over time intervals of $\approx$ one second in both the original data set and one created by averaging over ten frames. Between two and four such pixels typically appear in each time interval, but their locations changed as did their periods. The pixels are generally located at the edges of loop-like structures, where the field line geometry is likely to be simpler, but the periodicity is never sustained for longer than a few seconds. This is rather similar to the result of the investigation of \inlinecite{sin09} who found most oscillatory power at the edges of active-region loops. However, we are doubtful, on the basis of statistical tests and the fact that the oscillations are never long-lasting, whether these points have much significance.

An automated program was written in IDL and run in batch mode to analyze the emission in each of these pixels for every green-line image in the three time spans using wavelet routines (these are referred to as ``original" data). The temporal resolution for each sequence is thus 0.0256~s. Our program also analyzed the emission by taking averages of ten consecutive images, again using wavelet routines. For the time span covered by images 1001--6000, this gave a series of 500 images with time resolution 0.256~s and having a signal-to-noise ratio improved by a factor $\sqrt {10} = 3.162$ (``averaged" data).

\begin{figure}    
   \centerline{\hspace*{0.015\textwidth}
               \includegraphics[width=0.3\textwidth,angle=-90,clip=]{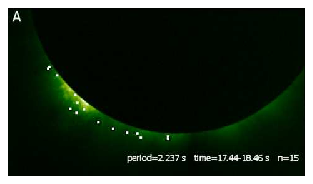}
               \vspace*{-0.03\textwidth}
               \includegraphics[width=0.3\textwidth,angle=-90,clip=]{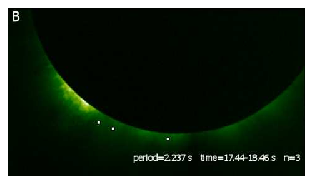}
              }
              \vspace{0.02\textwidth}
%
   \centerline{\hspace*{0.015\textwidth}
               \includegraphics[width=0.3\textwidth,angle=-90,clip=]{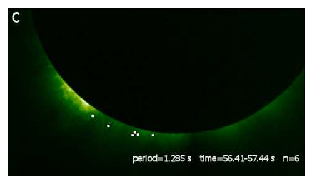}
               \vspace*{-0.03\textwidth}
               \includegraphics[width=0.3\textwidth,angle=-90,clip=]{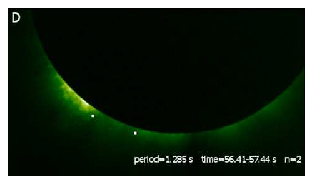}
              }
              \vspace{0.02\textwidth}


       \centerline{\hspace*{0.015\textwidth}
               \includegraphics[width=0.3\textwidth,angle=-90,clip=]{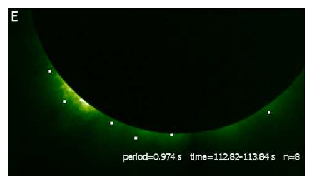}
               \vspace*{-0.03\textwidth}
               \includegraphics[width=0.3\textwidth,angle=-90,clip=]{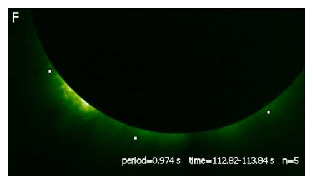}
              }
           \vspace{0.05\textwidth}

\caption{Locations of pixels (white dots, enlarged for clarity) showing significant periodicity ($>99.9$\%) for specific oscillation periods (2.24~seconds, 1.28~seconds, 0.97~seconds) and specific time spans assuming white noise (panels A, C, E) and global wavelet spectrum (panels B, D, F) as background spectrum. The original data set used has the full number (5000) of images.}
\label{period_analysis}
\end{figure}

Choosing particular oscillation periods [$P$], ranging from 0.05~seconds to 13.6~seconds (frequencies 19~Hz to 0.074~Hz), the program searched through the images in the original and averaged data sets for pixels showing enhanced wavelet power with $> 99.9$\% significance, counting the number of pixels and identifying them by their coordinates. The number of pixels with enhanced power at the 99.9\% significance level varied with period and time span, ranging from zero to only about four. Only a small fraction of these pixels in the original data were also evident in the averaged data. Those showing in both the original and averaged data sets were considered candidates for locations of possible real coronal oscillations. Although there is no obvious consistent pattern to their locations, there is a slight tendency for the pixels to cluster at any given moment around the edges of certain loop-like structures, but not all. This could be of significance since such locations are where the magnetic field structure is simplest -- nearer the photosphere there are likely to be more complex magnetic field patterns with overlapping wave motions if these are present.  Figure~\ref{period_analysis} shows the positions of these pixels for periods $P = 0.97$~seconds, 1.28~seconds, and 2.24~seconds. These pixels were selected through their enhanced power by comparison with some reference noise background. This background was assumed to be white noise, {\it i.e.} equal power at all frequencies (shown in the three left panels) or a  global wavelet spectrum (three right panels). As can be seen, the number of pixels with significant power drops considerably for the more restrictive global wavelet spectrum. The inconsistent nature of these points, changing with position and time, leads us to suspect that none indicates a solar origin.






\section{Conclusions and Future Directions}\label{Section 5}

Our analysis of such a huge data set ($476 \times 376$ pixels per frame, and 6500 frames in the green-line camera) from the SECIS experiment during the 21~June 2001 total solar eclipse, cannot be considered absolutely complete or the results final, but it does make clear that persistent periodicities in locations with significant signal in the green-line channel are not obviously present with any statistical significance. A first analysis (Section~\ref{first_anal}) revealed evidence of a few locations showing enhanced Fourier power, but they are isolated ({\it i.e.} neighbouring pixels do not show obvious periodicity) and the periodicities are not sustained for more than a few seconds for peaks having $> 4\sigma$ significance. A few pixels around the NOAA 9511/9513 complex show a strong periodic behaviour, with periods of about nine or ten~s, for about three cycles before being occulted by the Moon. The significance of the apparent periodicity cannot be properly assessed as it is sustained for such a short time.

A more comprehensive analysis (Section~\ref{comp_anal}) of all pixels having a signal of $\geqslant 10$~DN (dark current level is $\approx 2$~DN) -- a total of $10714$ locations -- similarly does not show up any significant periodic tendency that is sustained for more than one second. There are, however, many pixels having periodic tendencies over time intervals of $\approx$ one second in both the original data set and one created by averaging over ten frames. Between two and four such pixels typically appear in each time interval, but their locations changed as did their periods. The pixels are generally located at the edges of loop-like structures, where the field line geometry is likely to be simpler, but the periodicity is never sustained for longer than a few seconds. This is rather similar to the result of the investigation of \inlinecite{sin09} who found most oscillatory power at the edges of active region loops. However, we are doubtful, on the basis of statistical tests and the fact that the oscillations are never long-lasting, whether these points have much significance.

As has been pointed out by \inlinecite{pas09b} and others, ground-based observations, particularly during eclipses, have the advantage over space observations of the corona in that very high cadence observations are possible, whereas telemetry considerations limit image rates from spacecraft instruments. The extreme ultraviolet instruments such as EIT on SOHO, TRACE, and STEREO form images with a cadence that is generally one per several tens of seconds or minutes. The {\it Solar Dynamics Observatory}/AIA instrument is able to form images in ten wavelengths in about ten~s or better, but even this cadence is much less than the frequencies of fast-mode and slow-mode MHD waves that are considered theoretically possible for coronal heating, {\it viz.} $\approx$ one~Hz \cite{por94}.

Our cameras have ten-bit effective capacity and the optical set-up in the 1999 and 2001 eclipses resulted in a spatial resolution of four arcsec pixel$^{-1}$. Although considered state-of-the-art when manufactured in 1997, better quality cameras are now available. The latest results from Pasachoff's group \cite{pas02,rus00} for the 1999 eclipse was with 16-bit cameras and a larger telescope giving a theoretical two arcsec pixel$^{-1}$ resolution. Using Monte Carlo techniques, they found some positive evidence for 0.7--1~Hz oscillations. The \inlinecite{sin09} results for observations during the 2006 eclipse also used a larger telescope giving a theoretical resolution of 0.8~arcsec pixel$^{-1}$, but they were forced to bin their data in $2\times 2$ or $4\times 4$ pixel areas. They obtained short-exposure (100~ms or 300~ms) images with a cadence of 0.3~Hz and 0.5~Hz. The filters appear to be comparable in wavelength discrimination with the one we used in this work. Against this, our data had better temporal cadence (39 Hz) and also, with excellent seeing in nearly all our frames, fine loops are clearly visible if not completely resolved. The limitation imposed by the  ten-bit capacity of the cameras was also largely overcome by our choice of exposure times, with no pixels being over-exposed and nearly 40\% of available pixels having DN values of ten or more, so available for analysis.

It is possible that the spatial resolution of all eclipse instruments, even the most recent, may prove to be inadequate to resolve any oscillations that may be present. The theoretical results of \inlinecite{por94} imply that heating by MHD waves occurs in single loop structures, and it is still unclear whether such structures have ever been completely resolved. Moreover, any region showing a complex of loops such as the active regions observed in our data might show numerous oscillations in each line-of-sight as all the loops in the green-line are almost certainly optically thin. Active regions are therefore likely to be too complex to show recognizable intensity oscillations, and only simple structures such as loops in the outer parts of active regions or isolated structures are likely to show simple oscillations.

\inlinecite{pas02}  make this point concerning the alleged oscillations observed by \inlinecite{cow99} as their resolution was 18~arcsec, several times the width of loops observed by TRACE, which despite its one~arcsec resolution still does not appear to be resolving some structures.  Images taken during the 1990 eclipse with the Canada-France-Hawaii 3.6-m telescope, the largest ever used during a total eclipse \cite{kou93}, similarly show structures with dimensions $\approx$ one~arcsec. The widths of the finest coronal loop structures in the 2001 and other eclipses appear to be less than one~arcsec, as revealed by the image processing algorithms applied by Druckm\"uller and colleagues \cite{dru06,pas09a} to high-quality images. Any spatial movement of loop structures arising from oscillations appear to be beyond the capability of any instrument to resolve, as has been shown by \inlinecite{tom09}. For the velocity waves observed with their CoMP instrument, having amplitudes of 0.5~km~s$^{-1}$ and periods of around five~minutes, the spatial amplitude corresponds to only about 50~km, far smaller than is possible for any instrument to resolve. If shorter-period waves do exist and have similar velocity amplitudes, the spatial amplitudes will be correspondingly lower.

There will be no total solar eclipses for a few years convenient for observations like those discussed here, the next few being largely over sea or at least remote from Europe, and this has dictated the future directions for our research. Plans are underway to use the mountain-top coronagraph at Lomnick\'y \'St\'it, consisting of a pair of 200~mm $f/15$ telescopes \cite{amb10}.  One of the telescopes has already been used for tests in 2009 with the SECIS cameras described here. Only weak signals were apparent as the corona was so faint during the extended solar minimum over this period, but we are optimistic for future work when solar activity increases. Improved cameras are being considered, with 1~Mpixel image formats, as is a set-up with a pair of narrow-band filters having central wavelengths to the red and blue of the green-line wavelength, allowing the possibility of a ``dopplerometer" to look for velocity shifts, as with the CoMP instrument. Tests are planned for the coming year or so.

%
\begin{acks}
We thank the help and support given by Habatwa Mweene and his colleagues at the State University of Zambia, Lusaka, during the expedition, and the SOHO/CDS team for their support of our observations. Roger Thomas (NASA Goddard Space Flight Center, Greenbelt, MD, USA) is thanked for his contribution to the measurements of the Fe~{\sc xiv} green line wavelength and Francisco Diego and Jan Rybak for help with measuring our green-line filter characteristics. Josephine Chan's contribution to the analysis software helped a great deal in the early analysis of the data. Many of the images in this work were analyzed with Dominic Zarro's useful mapping software. PR acknowledges financial support from KBN grant number 2PO3D00551, with insurance costs generously paid for by the Polish insurance company PZU S.A. FPK is grateful to AWE Aldermaston for the award of a William Penney Fellowship. This work was supported by the Royal Society and Leverhulme Trust.

\end{acks}

%
%
\bibliographystyle{spr-mp-sola}
\bibliography{Search_for_rapid_changes}



\end{article}
\end{document}